\documentclass[nature,superscriptaddress,reprint,floatfix]{revtex4-1}

\usepackage[plainpages=false, colorlinks=true, anchorcolor=blue, linkcolor=blue, citecolor=blue, bookmarks=false]{hyperref}
\usepackage{amsmath, amsthm, amssymb}
\usepackage{graphicx}

\usepackage{soul}
\usepackage{color}
\usepackage{dcolumn}
\usepackage{bm}
\usepackage[mathscr]{euscript}
\usepackage{mathrsfs}
\usepackage{amsbsy}

\begin{document}

\title{Strong suppression of heat conduction in a laboratory replica of galaxy-cluster turbulent plasmas}


\author{J. Meinecke}
\affiliation{Department of Physics, University of Oxford, Parks Road, Oxford OX1 3PU, UK}
\author{P. Tzeferacos}
\affiliation{Department of Physics, University of Oxford, Parks Road, Oxford OX1 3PU, UK}
\affiliation{Department of Astronomy and Astrophysics, University of Chicago, 5640 S. Ellis Ave, Chicago, IL 60637, USA}
\affiliation{Flash Center for Computational Science, Department of Physics and Astronomy, University of Rochester, 206 Bausch \& Lomb Hall, Rochester, NY 14627, USA}
\affiliation{Laboratory for Laser Energetics, University of Rochester, 250 E. River Rd, Rochester, NY 14623, USA}
\author{J. S. Ross}
\affiliation{Lawrence Livermore National Laboratory, Livermore, CA 94550, USA}
\author{A. F. A. Bott}
\affiliation{Department of Physics, University of Oxford, Parks Road, Oxford OX1 3PU, UK}
\affiliation{Department of Astrophysical Sciences, University of Princeton, 4 Ivy Ln, Princeton, NJ 08544, USA}
\author{S. Feister}
\affiliation{Department of Astronomy and Astrophysics, University of Chicago, 5640 S. Ellis Ave, Chicago, IL 60637, USA}
\author{H.-S. Park}
\affiliation{Lawrence Livermore National Laboratory, Livermore, CA 94550, USA}
\author{A. R. Bell}
\affiliation{Department of Physics, University of Oxford, Parks Road, Oxford OX1 3PU, UK}
\author{R. Blandford}
\affiliation{Department of Physics, Stanford University, Stanford, CA 94309, USA}
\author{R. L. Berger}
\affiliation{Lawrence Livermore National Laboratory, Livermore, CA 94550, USA}
\author{R. Bingham}
\affiliation{Rutherford Appleton Laboratory, Chilton, Didcot OX11 0QX, UK}
\affiliation{Department of Physics, University of Strathclyde, Glasgow G4 0NG, UK}
\author{A. Casner}
\affiliation{CEA, DAM, DIF, 91297 Arpajon, France}
\author{L. E. Chen}
\affiliation{Department of Physics, University of Oxford, Parks Road, Oxford OX1 3PU, UK}
\author{J. Foster}
\affiliation{AWE, Aldermaston, Reading, West Berkshire RG7 4PR, UK}
\author{D. H. Froula}
\affiliation{Laboratory for Laser Energetics, University of Rochester, 250 E. River Rd, Rochester, NY 14623, USA}
\author{C. Goyon}
\affiliation{Lawrence Livermore National Laboratory, Livermore, CA 94550, USA}
\author{D. Kalantar} 
\affiliation{Lawrence Livermore National Laboratory, Livermore, CA 94550, USA}
\author{M. Koenig}
\affiliation{Laboratoire pour l'Utilisation de Lasers Intenses, UMR7605, CNRS CEA, Universit{\`e} Paris VI Ecole Polytechnique, 91128 Palaiseau Cedex, France}
\author{B. Lahmann}
\affiliation{Massachusetts Institute of Technology, Cambridge, MA 02139, USA}
\author{C.-K. Li}
\affiliation{Massachusetts Institute of Technology, Cambridge, MA 02139, USA}
\author{Y. Lu}
\affiliation{Flash Center for Computational Science, Department of Physics and Astronomy, University of Rochester, 206 Bausch \& Lomb Hall, Rochester, NY 14627, USA}
\author{C. A. J. Palmer}
\affiliation{Department of Physics, University of Oxford, Parks Road, Oxford OX1 3PU, UK}
\author{R. Petrasso}
\affiliation{Massachusetts Institute of Technology, Cambridge, MA 02139, USA}
\author{H. Poole}
\affiliation{Department of Physics, University of Oxford, Parks Road, Oxford OX1 3PU, UK}
\author{B. Remington}
\affiliation{Lawrence Livermore National Laboratory, Livermore, CA 94550, USA}
\author{B. Reville}
\affiliation{Max-Planck-Institut f\"ur Kernphysik, Postfach 10 39 80, 69029 Heidelberg, Germany}
\author{A. Reyes}
\affiliation{Flash Center for Computational Science, Department of Physics and Astronomy, University of Rochester, 206 Bausch \& Lomb Hall, Rochester, NY 14627, USA}
\author{A. Rigby}
\affiliation{Department of Physics, University of Oxford, Parks Road, Oxford OX1 3PU, UK}
\author{D. Ryu}
\affiliation{Department of Physics, UNIST, Ulsan 689-798, Korea}
\author{G. Swadling}
\affiliation{Lawrence Livermore National Laboratory, Livermore, CA 94550, USA}
\author{A. Zylstra}
\affiliation{Los Alamos National Laboratory, Los Alamos, NM 87545, USA}
\affiliation{Lawrence Livermore National Laboratory, Livermore, CA 94550, USA}
\author{F. Miniati}
\affiliation{Department of Physics, University of Oxford, Parks Road, Oxford OX1 3PU, UK}
\author{S. Sarkar}
\affiliation{Department of Physics, University of Oxford, Parks Road, Oxford OX1 3PU, UK}
\author{A. A. Schekochihin}
\affiliation{Department of Physics, University of Oxford, Parks Road, Oxford OX1 3PU, UK}
\author{D. Q. Lamb}
\affiliation{Department of Astronomy and Astrophysics, University of Chicago, 5640 S. Ellis Ave, Chicago, IL 60637, USA}
\author{G. Gregori}
\affiliation{Department of Physics, University of Oxford, Parks Road, Oxford OX1 3PU, UK}
\affiliation{Department of Astronomy and Astrophysics, University of Chicago, 5640 S. Ellis Ave, Chicago, IL 60637, USA}

\begin{abstract}
{\bf Galaxy clusters are filled with hot, diffuse X-ray emitting plasma, with
a stochastically tangled magnetic field \color{black}{whose energy is }\color{black} close to equipartition with \color{black}{the energy of }\color{black} the turbulent motions \cite{zweibel1997, Vacca}.
In the cluster cores, the temperatures remain anomalously high compared to what might be expected considering that the radiative cooling time is short relative to the Hubble time \cite{cowie1977,fabian1994}. While feedback from the central active galactic nuclei (AGN) \cite{fabian2012,birzan2012,churazov2000} is believed to provide most of the heating, there has been a long debate as to whether conduction of heat from the bulk to the core \color{black}{can }\color{black} help the core to reach the observed temperatures \cite{narayan2001,ruszkowski2002,kunz2011}\color{black}{, given the presence of tangled magnetic fields}\color{black}. Interestingly, evidence of very sharp
temperature gradients in structures like cold fronts implies a high degree of suppression of thermal conduction
\cite{markevitch2007}.
To address the problem of thermal conduction in a magnetized and turbulent plasma, we have created a replica of such a system
in a laser laboratory experiment. 
Our data shows a reduction of local heat transport by two orders of magnitude or more, leading to \color{black}{strong temperature variations on small spatial scales}\color{black}, as \color{black}{is seen in cluster }\color{black} plasmas \cite{markevitch2003}.}
\end{abstract}

\maketitle

There are a number of possible mechanisms that can lead to a reduction from the classical (local) \color{black}{Spitzer }\color{black} conductivity: electrons getting stuck in local magnetic mirrors \cite{komarov2016}, alignment of local temperature gradients perpendicular to magnetic fields 
\cite{komarov14} and plasma instabilities such as whistler waves \cite{komarov2018,roberg2018}. For these processes to occur, the \color{black} electron's \color{black} Larmor radius must be small compared to \color{black} its \color{black} Coulomb mean free path, enabling microscale \color{black} changes in the electron heat transport \color{black} to alter the global properties of the plasma dynamics.  
Numerical simulations are \color{black} unable \color{black} to address these issues fully \cite{zuhone2012,smith2013}, as comprehensive modelling of all the physical processes present at different scales remains very challenging computationally.
Laboratory experiments \cite{gregori2015} can provide an alternative approach if they can achieve sufficiently weakly-collisional conditions \color{black} and sufficiently strong stochastic magnetic fields \color{black} for the heat transport to be significantly modified. 

Here we report such an experiment \color{black} using \color{black} the National Ignition Facility (NIF) laser at the Lawrence Livermore National Laboratory -- see Figure 1 for details of the experimental setup.  
The platform is similar to that already employed at other laser facilities (e.g., the Omega laser \cite{tzeferacos2018, bott2021time}) but with \color{black} $\sim$30 times more \color{black} energy delivered to the target. A highly turbulent plasma was created by ablating two doped plastic foils, each with 133 kJ of 351 nm light in 15 ns (see Figure 1).  The foils were separated by 8 mm and ablated on the back side to create colliding plasma flows at the centre (4 mm from each target). 
To help excite turbulence, plastic mesh grids with 300 $\mu \rm m$ apertures and 300 $\mu \rm m$ wires were placed 2 mm from each foil to disturb the two flows before they collided.  Three-dimensional simulations \color{black} carried out with \color{black} the FLASH code (validated on previous experiments at smaller laser facilities \cite{tzeferacos2017,bott2021time}) \color{black} were used to help design the platform and analyze \color{black} the experimental data.

\color{black} Several complementary \color{black} plasma diagnostics were deployed (see Methods).
With optical Thomson scattering, we measured an average electron density $n_e \approx 5\times 10^{20}$ cm$^{-3}$ in the collision region at $t=25$ ns after the start of the drive beams. The turbulent velocity in the plasma was determined via stimulated Brillouin scattering. By recording the
wavelength shift of the back-scattered light from a probe beam, we infer $v_{\rm turb}\approx 200$ km/s.
\color{black} The \color{black}self-generated magnetic fields were estimated from proton deflectometry to be  $B_{\text{RMS}} \approx 0.8 (B_{\rm path, \|}/25 \, \mathrm{kG \, cm}) (\ell_B/100 \, \mu \mathrm{m})^{1/2}(\ell_n/2 \, \mathrm{mm})^{1/2}$ MG (here $B_{\rm path, \|}$ is the \color{black}{one } \color{black} component of the magnetic field \color{black}{that we measure}\color{black}), 
where we took the proton path-length $\ell_n$ from self-emission X-ray images \cite{tzeferacos2018,bott2021time},
and the magnetic field correlation length, $\ell_B$, corresponds to a wavelength ${\sim}4 \ell_B \approx 400$ $\mu$m of the same order as the grid periodicity. Using the same diagnostic, we found the maximum field to be
$B_{\text{max}} \gtrsim 3$ MG.

By comparing soft X-ray emission images in two wavelength bands determined by filtering the broadband emission with either 6.56 $\mu$m polyimide or 2.36 $\mu$m vanadium (see Figure 2), two-dimensional maps of 
{\color{black} $\langle T_{\rm e}\rangle_X$, a measure} of the electron temperature averaged along the line of sight, were obtained with ${\sim}$50 $\mu$m spatial resolution and $\sim$100 ps temporal resolution \color{black}
(see Supplementary Information for details and validation with synthetic FLASH data). 
This temperature diagnostic \color{black} closely resembles \color{black} what has recently been used for galaxy-cluster X-ray analysis \cite{churazov2016}. 

At \color{black}{$t\gtrsim 23$ ns }\color{black} after the start of the drive beams, we measured an average 
temperature measure {$\langle T_{\rm e} \rangle_X  \approx 1.1$ keV }\color{black} over a 2 mm $\times$ 1 mm region: see Figure 3.  The \color{black} profile of the \color{black} \color{black}{$\langle T_{\rm e} \rangle_X$ }\color{black} map \color{black} is highly structured, \color{black}  with the magnitude of local perturbations in $\langle T_{\rm e} \rangle_X$ exceeding ${\sim}1$ keV by $t = 25$ ns. \color{black}
The spikiness of the $\langle T_{\rm e} \rangle_X$ profile is less pronounced at earlier times,   
\color{black} and very \color{black} different from the temperature maps that we obtained in previous experiments at \color{black} less energetic \color{black} laser facilities \cite{tzeferacos2018}, which showed a nearly uniform temperature distribution.

Taking the measured RMS magnetic field, we find that the electrons in the interaction region are strongly magnetized (see Supplementary Information), viz.,~$r_g/\lambda_e \sim 0.08$, where $r_g$ is their gyroradius and \color{black} $\lambda_e \approx 0.1$ $\mu$m their  mean free path. 
Note $\lambda_e \lesssim \ell_T$,
where $\ell_T \lesssim 50$ $\mu$m is the thermal gradient \color{black} length scale \color{black} (limited by the spatial resolution of the diagnostics),
implies that ordinary (Spitzer) conductivity is somewhat modified by non-local effects \cite{bell1981}.
Non-local thermal conduction has been previously seen in laboratory experiments \cite{gregori2004} and results in a smoothing of the heat front -- increasing its width by a factor ${\sim}5$ for $\lambda_e \sim \ell_T$ \cite{komarov2018}. However, in the presence of magnetic fields, electron conduction perpendicular to the field lines is further reduced by up to a factor ${\sim}\lambda_e/r_g$.
The parallel heat conduction is also quenched by a factor of $3/\beta_e$
due to the development of the whistler instability \cite{komarov2018}.

Evidence of significant reduction of heat conduction in our NIF experiment is illustrated by the temperature maps in Figure 3.
The image shows localized hot patches with scale length $\ell_T \lesssim 50$ $\mu$m. 
For normal conduction, we should expect the time for the temperature gradients associated with the hot spots to disappear to be $t_{\rm cond} \sim k_B n_e \ell_T^2/\kappa_S \sim 8\times 10^{-12}$ s, where $\kappa_S$ is the Spitzer thermal conductivity. 
Instead, as we do see temperature structures, they must have existed for a dynamical time 
$t_{\rm age} \sim L/c_s \sim 3\times 10^{-9}$ s, where $L\sim 1$ mm is the spatial extent of the interaction region and $c_s$ is the sound speed.
This implies a reduction in the effective conductivity ($\kappa$) by a factor $(\kappa/\kappa_S)^{-1} \sim (t_{\rm age}/t_{\rm cond}) \gtrsim 100$--$200$ between $t=23$--25 ns.

The notion that the reduction in heat conduction is due to electrons being magnetised can be further strengthened if we compare NIF results with a turbulent plasma where $r_g/\lambda_e \gtrsim 1$, which was achieved in our previous experiments at the Omega laser facility, shown in Figure 3c \cite{tzeferacos2018}.
While the turbulence itself exhibits a similar structure, the temperature map in this case is significantly more homogeneous, with $\ell_T \gg \lambda_e, \, \ell_B$, and the same considerations as above lead to $(\kappa/\kappa_S)^{-1}\sim 1$, as expected for normal conduction. 

Our analysis is supported by numerical simulations of the Omega and NIF experiments performed with the FLASH code.  \color{black} For both experiments, we generated X-ray images from the simulation outputs and synthetic temperature maps (Figures 3g,h,i - see Supplementary Information for details). For Omega, the code, run with Spitzer's conductivity, accurately reproduces the experimental $\langle T_{\rm e} \rangle_X$ maps. For NIF, \color{black} we compared simulation results for the cases when \color{black}{Spitzer }\color{black} thermal conduction was turned off \color{black} to \color{black} those with it included.
 \color{black}{We find that a highly structured $\langle T_{\rm e} \rangle_X$ \color{black} profile \color{black} is only obtained in the conduction-off simulations. }\color{black}

More quantitatively, the distribution of $\langle T_{\rm e} \rangle_X$ fluctuations in the interaction region agrees between the experimental data and synthetic predictions \color{black}{from the FLASH conduction-off simulations (Figure 4a). 
\color{black}
Even though FLASH simulations lack the kinetic physics likely responsible for the suppression of conductivity in the experiment, this agreement is unsurprising for the following reason. In the experiment, because thermal conductivity is suppressed, mixing of temperature perturbations by turbulent motions predominates over thermal diffusion for the smallest temperature fluctuations that are resolved by our diagnostic. The conduction-off simulations have a similar resolution (25 $\mu$m) and a very small artificial viscosity (see Supplementary Information), so they capture the turbulent motions and the resulting temperature fluctuations on the same scales as the experimental diagnostic.

\color{black}
Further analysis of the simulations suggests a plausible candidate for the mechanism that gives rise to the temperature inhomogeneities (that are then mixed by turbulent motions): a radiative-cooling instability that acts on the initial temperature perturbations in the interaction region arising from asymmetries in the jets' collision. The instability is a result of line cooling due to the dopants in the plasma, which causes the cooling function for the plasma to be a strongly decreasing function of the temperature for much of the relevant range of temperatures (see Supplementary Information). \color{black} As a result, regions in the simulated plasma where $T_e \lesssim 400$ eV and $700$ eV $\lesssim T_e \lesssim 900$ eV experience significant radiative cooling on a timescale that is comparable with the eddy-turnover time of the turbulence (see Figure \color{black} 4b).
Also, approximate pressure balance across the whole interaction region (see Supplementary Information) implies that the regions that experience significant radiative cooling must undergo compression, while those that do not must expand \citep{TS1974,H81}, enhancing the impact of the cooling instability across the whole plasma. The instability cannot operate in the conduction-on simulations because the initial temperature fluctuations in the plasma are rapidly suppressed by efficient heat transport. 
}\color{black}

Our results provide the first direct experimental evidence of suppression of heat conduction in a turbulent magnetized plasma for conditions that resemble those of galaxy clusters, and more generally in high-$\beta$ turbulent plasma with weakly collisional and magnetised electrons. They suggest that \color{black} the \color{black}
effective electron conductivity is about two orders of magnitude below that predicted by Spitzer's theory, and so, e.g., in cluster cores, conduction is unlikely to be \color{black} able to lower \color{black} the AGN feedback requirements, \color{black} which requires  a conductivity larger than  $\approx \kappa_S/5$ \color{black} \cite{narayan2001,ruszkowski2002}.
Since both the experiment and the cluster plasma are in the regime where the electron Larmor radius is smaller than the mean free path, such strong modification in the heat transport points to plasma micro-instabilities as the main culprit for the reduction of classical Spitzer's conductivity \cite{komarov2018,roberg2018}. Precisely how to calculate the
effective conductivity in such a plasma is
currently unknown, either theoretically or numerically. Our NIF measurements thus provide
an experimental benchmark for the development of future multi-scale models of heat transport in turbulent and magnetized plasmas.

\subsection*{Materials and Methods}
Optical Thomson scattering (OTS) from collective electron plasma wave (EPW) oscillations was used to characterise the electron density in the interaction region of the two counter-propagating turbulent flows \cite{evans1969}.  Four laser beams were focused at the interaction region with a spot radius of 600 $\mu \rm m$, forming a cylindrical collection volume of diameter 50 $\mu \rm m$ and length of 1.2 mm (see Figure 1).  From a scattering angle of $40^{\text{o}}$, light was collected by a spectrometer, dispersed by a grating, and streaked onto a camera to measure the temporal evolution of the electron density. The peak position of the EPW is mainly determined by the plasma frequency (via the plasma dispersion relation) \cite{evans1969}. Since scattering occurs over a region which is much longer than the laser wavelength,
in a turbulent plasma this implies that different electron densities are probed simultaneously, and so the observed broadening of the EPW is a measurement of that range of densities.

Similar to OTS is the stimulated Brillouin scattering (SBS) diagnostic,
which measures the light scattered back in the laser direction in a narrow wavelength range near the laser frequency. 
The SBS instability results from the resonant coupling between the probe laser light, the scattered light and an ion-acoustic wave \cite{rose1994}.
To drive this instability, we use four NIF beams delivering 0.5 TW in a 100 $\mu \rm m$ spot, probing different points in the interaction region. The back-scattered light (mostly occurring near the peak of the density profile) shows a frequency shift $\Delta\lambda_{\text{SBS}} = \Delta\lambda_{\text{iaw}} + \Delta \lambda_{\text{flow}}$, where $\Delta\lambda_{\text{iaw}}$ is associated with the coupling of the probe light with ion-acoustic waves, while $\Delta \lambda_{\text{flow}}$ is the Doppler contribution from the plasma 
flow moving in the direction of the laser beam.
If the plasma temperature is known (from the temperature maps shown in the main text), then $\Delta\lambda_{\text{iaw}}$ can be estimated, and the SBS diagnostic measures the component of the flow moving along the probe beam, which is nearly perpendicular to the axis connecting the two grids (see Figure 1).  
Since the velocities near the centre of the interaction region are chaotic (as confirmed by the  FLASH simulations), the SBS measurement is indicative of the turbulent velocities achieved in the experiment. 

The turbulence-amplified magnetic fields were inferred from the angular deflections of energetic protons as they traversed the interaction region. These were created by illuminating a D$^3$He capsule
\color{black} with \color{black} 60 additional laser beams delivering a total of 48 kJ in a 900 ps pulse. Fusion reactions in the imploding capsule generate nearly monoenergetic 15 MeV protons. 
Unlike previous experiments where protons of these energies only acquired small deflections,  \color{black} proton deflections in the present experiment \color{black} are large, with the trajectory of each individual proton crossing over those of other protons before reaching the detector. \color{black} This \color{black} prevents the application of previous analysis techniques capable of reconstructing two-dimensional maps of the path-integrated magnetic field \cite{bott2017}.

In order to characterize the magnetic field, we instead measured the proton deflections by placing a slit in the path of the protons to limit the
size of the beam as it entered the plasma. 
Qualitatively, the significant displacement of protons from their projected positions in the absence of any deflection is consistent with strong magnetic fields. More quantitatively, we measured the mean, the root mean square (RMS), and the maximum displacement of the proton flux distribution from the slit's central position.  This allows us to estimate the corresponding mean, RMS and maximum values of the component of the path-integrated magnetic field parallel to the slit's orientation ($B_{\rm path, \|} \equiv \int \mathrm{d} s \, B_{\|}$). 
We also \color{black} analyzed \color{black} inhomogeneities in the proton flux distribution and \color{black} found \color{black} that their scale \color{black} was \color{black} consistent with the correlation length of the magnetic field \color{black} being \color{black} $\ell_B \approx 100 \, \mu \mathrm{m}$. 
Further details of this analysis are provided in the Supplementary Information.

\subsubsection*{Acknowledgements}The research leading to these results received funding from the EPSRC (grant numbers EP/M022331/1 and EP/N014472/1), the U.S. DOE under Contract No. B591485 to LLNL, Field Work Proposal No. 57789 to ANL and DE-SC0016566 to the University of Chicago, DE-NA0003868 to the Massachusetts Institute of Technology, DE-NA0001808, 89233118CNA000010, and 89233119CNA000063 to General Atomics, subcontracts no. 
536203 and 630138 (LANL) and B632670 (LLNL) to the Flash Center for Computational Science, and Cooperative Agreement DE-NA0003856 to the Laboratory for Laser Energetics, University of Rochester. We acknowledge support from the NSF under grants PHY-1619573 and PHY-2033925. Awards of compute time were provided by the U.S. DOE ALCC program.
We acknowledge funding from grants 2016R1A5A1013277 and 2020R1A2C2102800 of the NRF of Korea.
The work of AAS was supported in part by the U.K. EPSRC Programme Grant EP/R034737/1. Support from AWE plc. and the STFC of the U.K. is also acknowledged.

\bibliography{Refs}

\newpage

\begin{widetext}
\newpage

\begin{figure}
\centering 
\includegraphics[width=6.in]{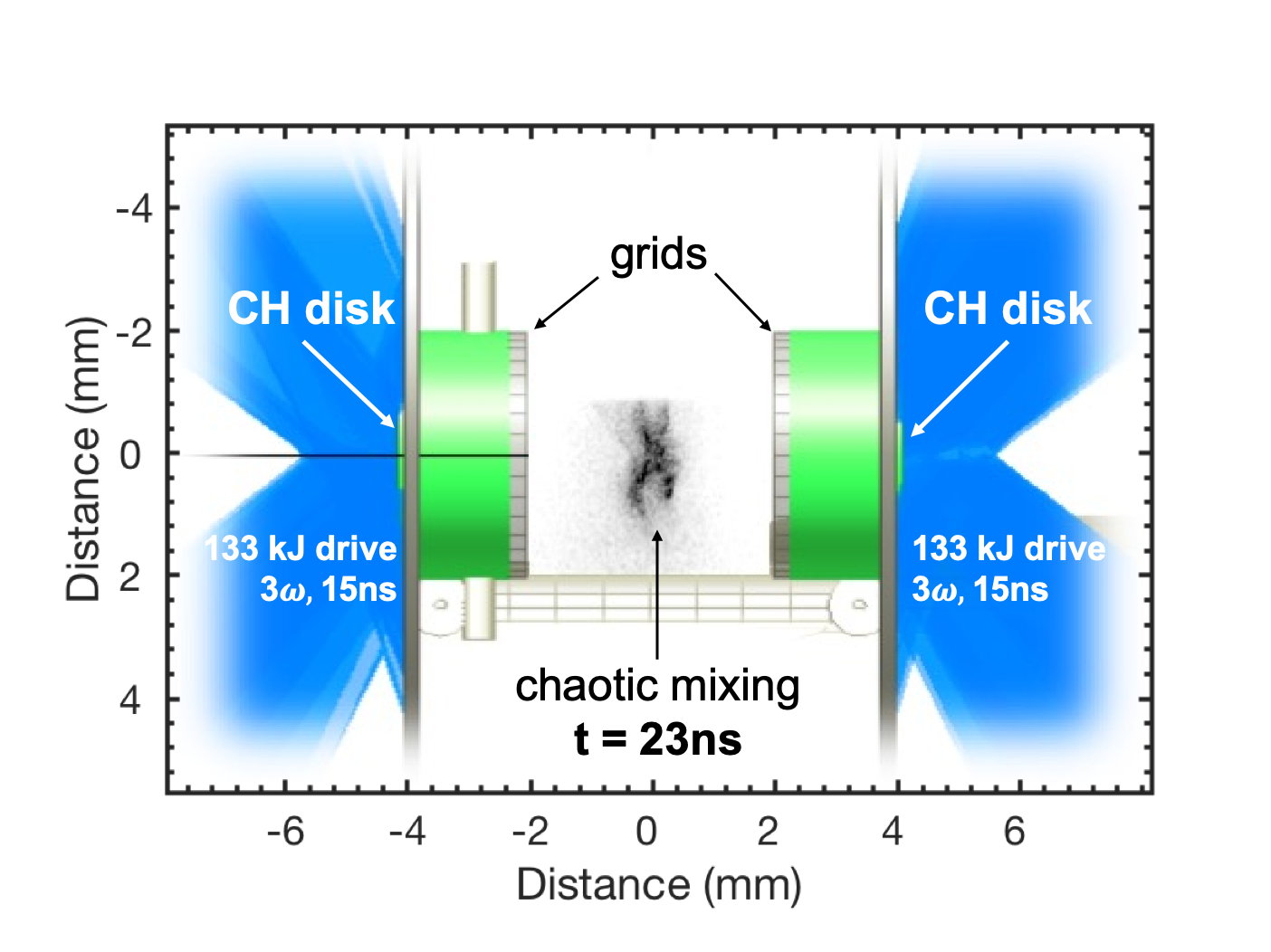}
\caption{{\bf Experimental configuration.}  Two CH disks are separated by 8 mm and  ablated with 96 frequency-tripled (351 nm wavelength) laser beams with a 1 mm spot diameter.  The total laser illumination onto each disk is 133 $\pm$ 10 kJ in a 15 ns square pulse length. Turbulence is generated with plastic grid obstacles  (300 $\mu$m aperture, 300 $\mu$m wires) located 2 mm away from each disk by disturbing both collimated flows before interpenetration and mixing.  A gated X-ray detector (GXD) observes the chaotic mixing of the flows (shown here at 23 ns after beginning of the laser drive), while proton radiography characterises the self-generated magnetic fields.   A 860 $\mu$m diameter capsule filled with 6 atm of D$_2$ gas  and 12 atm of $^3$He, is located 18 mm from the midpoint of the two disks and ablated with 60 frequency-tripled laser beams with a 200 $\mu$m spot that deliver 43 $\pm$ 5 kJ in a 900 ps square pulse length.  The implosion produces 14.7 MeV mono-energetic protons that penetrate the  interaction region and   are collected by a CR-39 nuclear track detector.  An optical Thomson-scattering (OTS) probe of frequency-tripled light was focused to the midpoint between both disks, consisting of four beams, totalling 12.8 kJ in a 8 ns square pulse, to measure average 
electron density. In some shots, the OTS probe was swapped for a full-aperture backscatter system (FABS) which measured the reflected light from four frequency-tripled lasers focused on the interaction region with 5.7 kJ in a 12.6 ns picket pulse and a 150 $\mu$m spot, to measure the turbulent velocity in the plasma.
}
\end{figure}

\begin{figure}
\includegraphics[width=6.in]{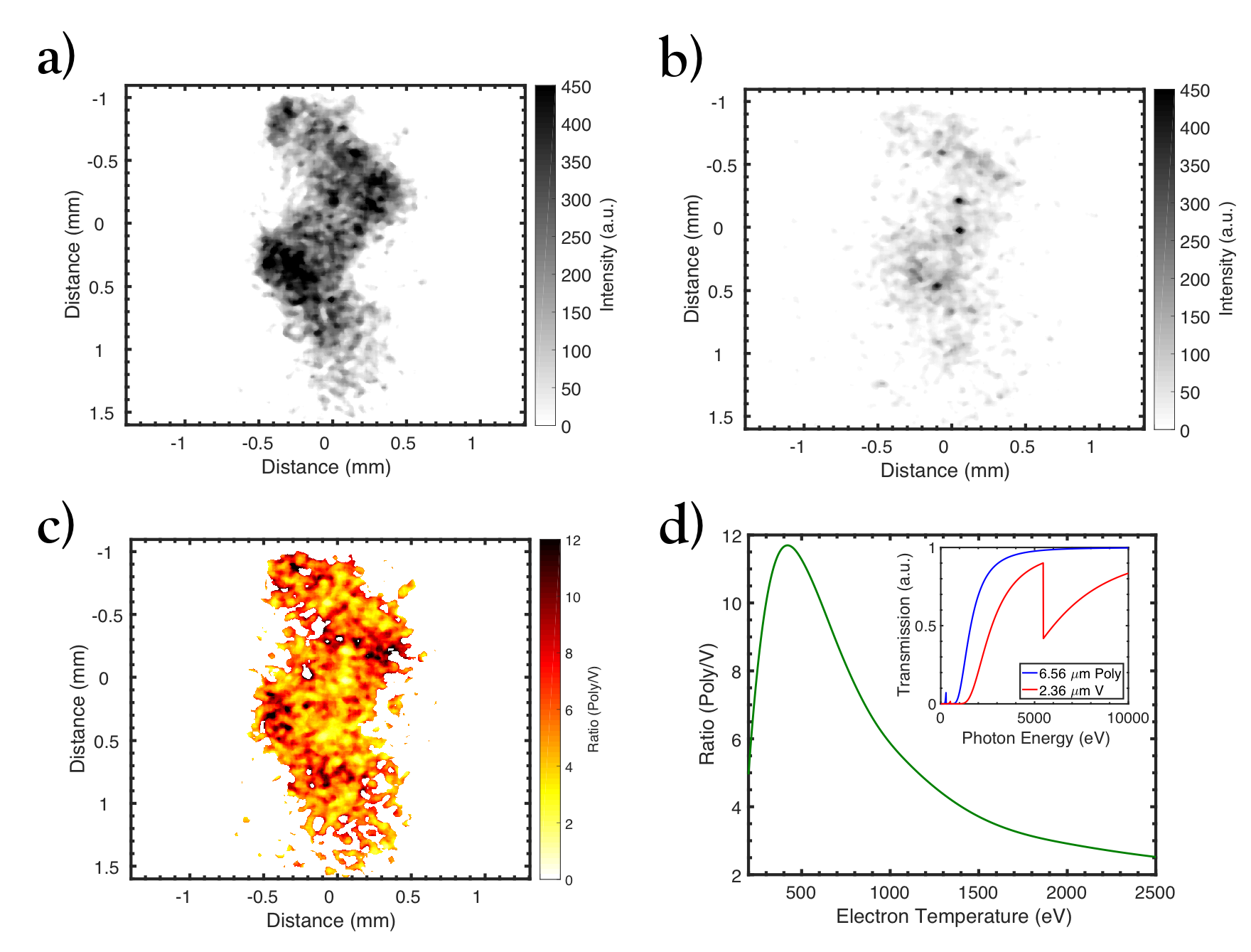}
\caption{{\bf X-ray emission images.} A two-pinhole array coupled to a GXD camera was used to image simultaneously the plasma interaction region where turbulence was generated, with a different filter in front of each aperture. {\bf (a)} Broadband X-ray emission filtered by 6.56 $\mu$m polyimide  and {\bf (b)} filtered by 2.36 $\mu$m vanadium foils \color{black} at 25 ns after the initial drive lasers fired. \color{black} The ratio of the X-ray emission in these two bands (polyimide/vanadium) is shown in panel {\bf (c)}. In producing the image, regions within $20 \%$ of the background intensity were excluded, and a 50 $\mu$m (corresponding to the \color{black} diameter of the pinholes\color{black}) smoothing was applied.
Panel {\bf d)} shows the relation between the X-ray intensity ratio of polyimide to vanadium to the electron temperature of the plasma. The measured X-ray signal depends on the filter transmission (inset), the GXD camera response, and the plasma emission, which is a function of density and temperature. Since the wavelength dependence of the plasma emission is (mostly) insensitive to the density, the X-ray signal strengths from two different energy bands is a strong function of the temperature. \color{black}
\color{black} When the temperature is \color{black} larger than 450 eV, a one-to-one correspondence between \color{black} the \color{black} X-ray intensity ratio and \color{black} the \color{black} plasma temperature is thus obtained.
}
\end{figure}

\begin{figure}
\includegraphics[width=7.in]{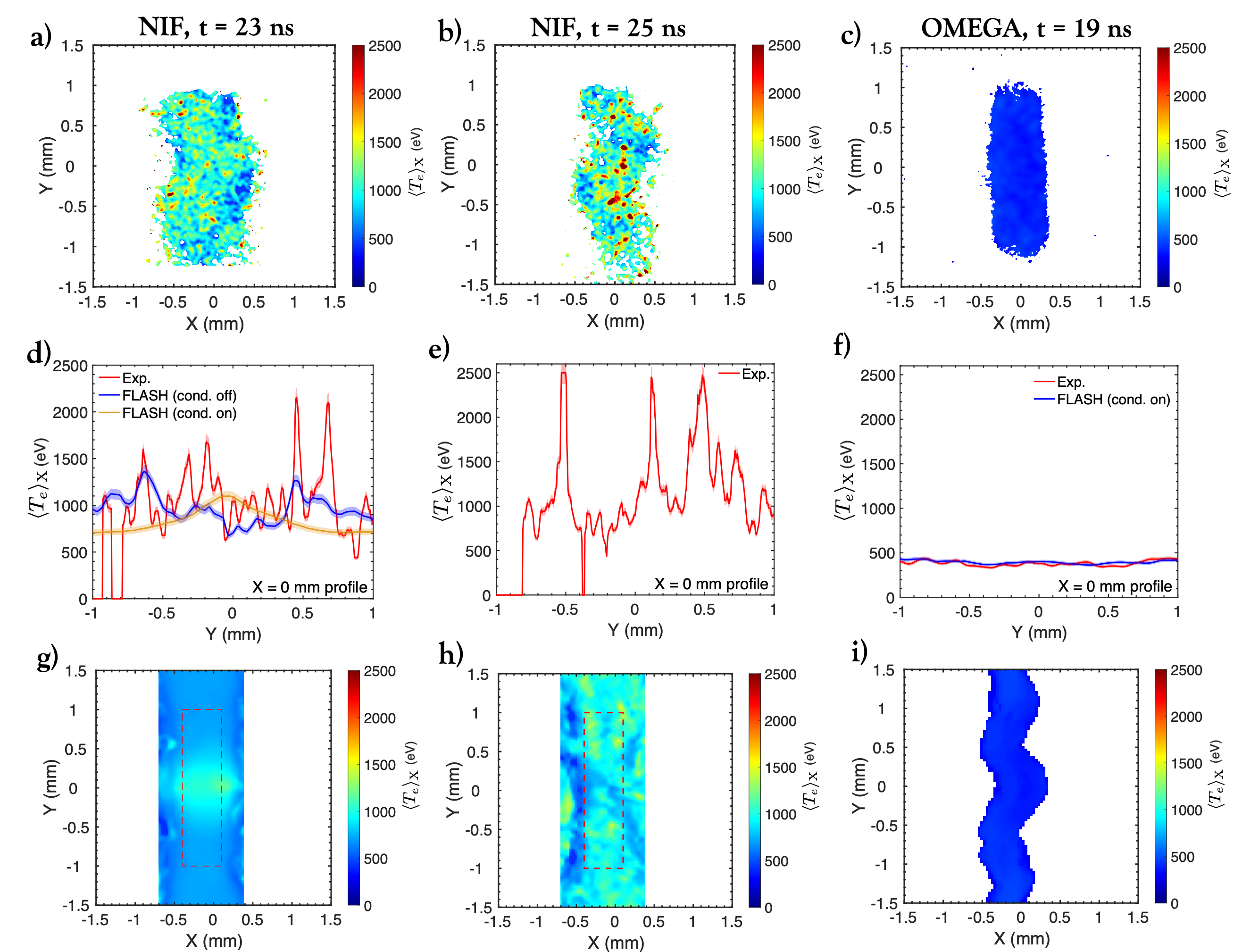}
\caption{{\bf Measured and simulated X-ray temperature maps.} {\bf (a)}
Map of the line-of-sight-averaged electron-temperature measure $\langle T_{\rm e} \rangle_X$ at $t=23$ ns after the start of the NIF laser drive. The maps are derived from the X-ray intensity ratio using the filter configuration described in Figure 2.   
{\bf (b)} Same as (a) but at $t=25$ ns.  
{\bf (c)} Same as (a), and for the same target as shown in Figure 1, but driven using the Omega laser at 19 ns after the initial drive lasers fired. In this case, the temperature map was constructed by comparing X-ray measurements taken with a 50 $\mu$m pinhole camera filtered by 4 $\mu$m Mylar plus 80 nm Al, and with 2 $\mu$m Mylar plus 40 nm Al (and assuming an average electron density of $10^{19}$ cm$^{-3}$). This diagnostic is sensitive to the temperature range $200$ eV $< \langle T_{\rm e} \rangle_X <$ 700 eV.  The temperature across the emitting surface was quite uniform on Omega.
Vertical temperature profiles taken at the $X=0$ mm position for each of the previous maps in (a), (b), and (c), are shown in {\bf (d)}, {\bf (e)}, and {\bf (f)} respectively, with a $5\%$ error band added to each line. \color{black} Equivalent temperature profiles from post-processed FLASH simulations are also depicted. 
{\bf (g)} Synthetic temperature map constructed by post-processing FLASH simulation results of the NIF experiment (using the X-ray intensity ratio and GXD response as in the NIF experi,ent) at $t = 23$ ns. The map shown in this panel was obtained for the case of Spitzer thermal conduction switched on.   
{\bf (h)} Same as panel (g) but with Spitzer thermal conduction switched off.
{\bf (i)} Same as panel (g), but for FLASH simulations of the Omega laser experiment at $t=19$ ns and using the same X-ray intensity ratio and GXD response as in the Omega experiment. The map shown in this panel was obtained for the case of Spitzer thermal conduction switched on. 
}
\end{figure}

\begin{figure}
\includegraphics[width=6.in]{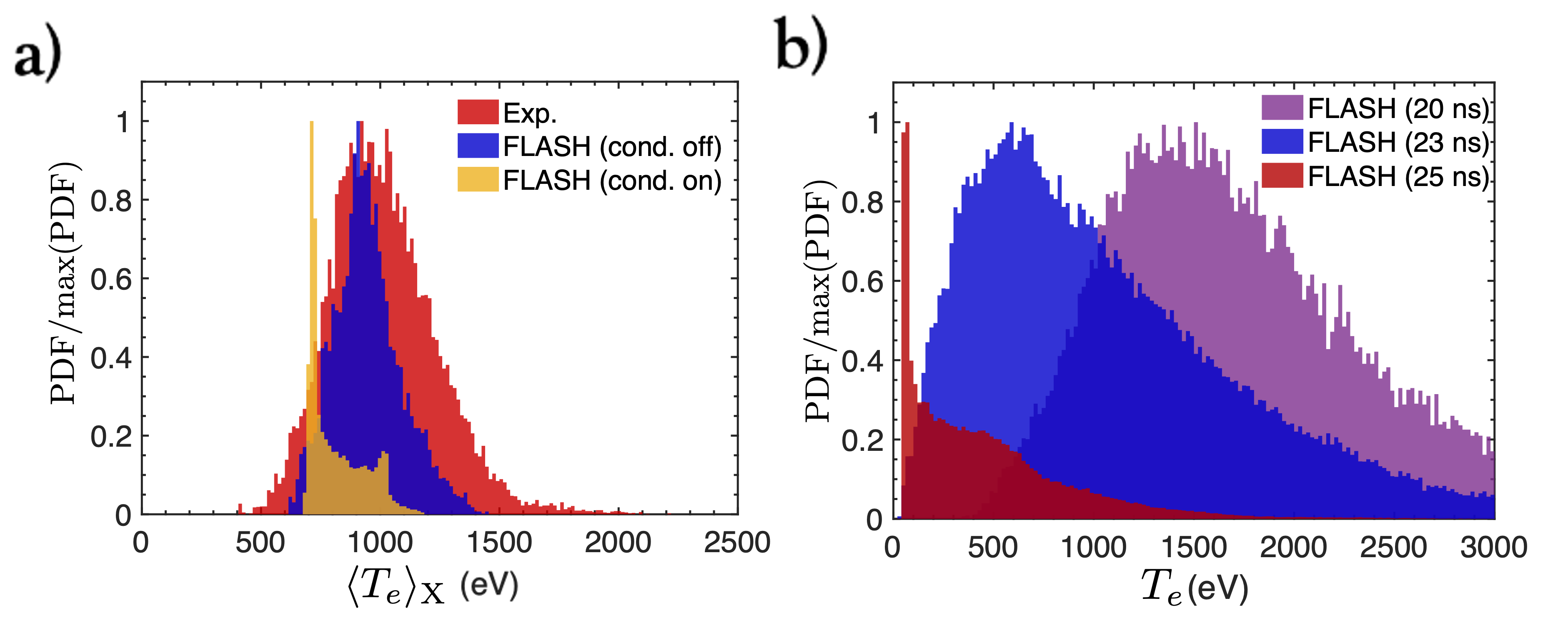}
\caption{{\bf Analysis of temperature fluctuations.}  
{\bf (a)} The distribution of $\langle T_{\rm e} \rangle_X$ from the experimental data (see Figure 3a) at $t = 23$ ns, \color{black} compared with \color{black} analogous distributions derived from the \color{black} conduction-on and conduction-off \color{black} FLASH simulations (see Figures 3g, h). In each case, the distribution is derived for an $0.5$ mm by $2$ mm area (demarcated by the dashed red lines in Figures 3g, h), and is normalised to its maximum value. {\bf (b)} Distribution of actual temperatures in the conduction-off FLASH simulations in a $0.5$ mm by $2$ mm$^2$ volume centred on the interaction region at $t = 20 \, \mathrm{ns}$ (just after the jet collision), $t = 23 \, \mathrm{ns}$ and $t = 25 \, \mathrm{ns}$. All points that are inside this volume but lie outside of the interaction region are excluded. 
}
\end{figure}

\end{widetext}

\end{document}


\maketitle

\section*{Calculation of plasma parameters}

The plasma was generated from the ablation of plastic foil targets (CH), manufactured by General Atomics, with trace amounts of homogeneously distributed dopants, listed in Table~\ref{table:values1}. Due to the composition of the doped targets, the effective ion charge seen by the free electrons was calculated for a multi-species plasma as follows:
\begin{equation}
Z_{\text{eff}} = \frac{1}{n_{\text{e}}} \sum_j Z_j^2 n_j,
\end{equation}
where the indices $j$ refer to each target element. From the quasineutrality condition, it follows that the electron density can be expressed as $n_{\text{e}} = \sum_j Z_j n_j$.  To simplify these equations, we can relate the density of each elemental component to that of hydrogen by setting $n_j = R_{j{\text{H}}}n_{\text{H}}$, where $R_{j{\text{H}}}$ is the ratio of a given element's concentration to the amount of hydrogen in each sample foil, e.g., $R_{{\text{OH}}} = 1.2/49.6$.  This reduces the effective ion charge equation to
\begin{equation}
Z_{\text{eff}} = \frac{\sum_j R_{j{\text{H}}}Z_j^2}{\sum_j R_{j{\text{H}}}Z_j}
\label{eq:zeff}
\end{equation}
for $j = $ H, C, O, N, Mn, Co, and $R_{{\text{HH}}}=1$.  To determine the effective ion charge as seen by the hydrogen atoms, Equation (\ref{eq:zeff}) is modified as $Z_{\text{eff,B}} = [(\sum_j R_{j{\text{H}}}Z_j^2)-Z_{\text{H}}^2]/ \sum_j R_{j{\text{H}}}Z_j$.   

Figure 3 in the main text shows a line-integrated electron temperature map over the 2 $\times$ 1 mm$^2$ interaction region. 
From the histogram of the temperature distribution, the mean
value corresponds to $T_{\text{e}} = T_{\text{i}}= 1.1$ keV, as shown in Table~\ref{table:values1} (assuming local thermodynamic equilibrium). The value at 10\% of the peak is  
$T_{\text{e}} = T_{\text{i}}= 1.6$ keV. 

The inflow and turbulent velocities were both measured directly with the time-resolved stimulated Brillouin scattering (SBS) diagnostic. 
The SBS diagnostic collected reflected light from four, frequency-tripled laser probe beams, collectively delivering 5.7 kJ of energy in a 12.6 ns picket pulse with a 150 $\mu$m spot size.  
Two of the SBS beams were pointed offset from the target chamber center (TCC), at 500 $\mu$m towards each of the foils, to measure the inflow velocity, determined from the start of the SBS signal. A third SBS beam, pointed at TCC, was used to determine the turbulent velocity, as discussed in the main text. 

\renewcommand{\arraystretch}{1.5}
\begin{table}[htbp]
\centering
\begin{tabular}{ c | c }
  {\bf Quantity} & {\bf Experimental Value} \\
  \hline
  
  Composition of target & 
  48.3\% C, 49.6\% H, 1.2\% O, \\  & 0.6\% N, 0.1\% Mn, 0.1\% Co \\
  Average atomic weight ($\langle M \rangle$)  &  6.7 \\
  Mean ion charge ($\langle Z \rangle$) & 3.6 \\
  Effective ion charge ($Z_{\text{eff}}$) & 5.7 \\
  Hydrogen effective ion charge ($Z_{\text{eff,B}}$) & 5.5 \\ 
  Electron temperature ($T_{\text{e}}$) & 1100 eV \\
  Ion temperature ($T_{\text{i}}$) & 1100 eV \\
  Electron number density ($n_{\text{e}}$) & $4.9 \times 10^{20}$ cm$^{-3}$ \\
  Turbulent velocity (${\text{v}}_{\text{turb}}$) & $2 \times 10^7$ cm s$^{-1}$ \\
  Outer scale ($L$) & 0.06 cm \\
  RMS magnetic field ($B_{\text{RMS}}$) & 0.8 MG \\
  Maximum magnetic field ($B_{\text{max}}$) & 3.0 MG \\
  Adiabatic index ($\gamma_{\text{I}}$) & $5/3$ \\

\end{tabular}
\caption{Summary of target characteristics and direct experimental measurements of plasma parameters at t $= 25$ ns after the start of the drive laser pulse. The outer scale, $L$, represents the distance between two neighboring grid aperture centers. Values are reported in CGS, except for the temperature that is given in eV.}
\label{table:values1}
\end{table}

Using the experimental values in Table~\ref{table:values1}, we calculated the relevant plasma parameters indicated in  Table~\ref{table:values2}.  Both the hydrogen-ion and electron-ion mean free paths are smaller than the system size ($L \approx 0.06$ cm), $\lambda_{\text{Hion}}, \lambda_{\text{e}} = 0.0013$ cm  $ \ll L$.

\renewcommand{\arraystretch}{2}
\begin{table}[htbp]
\centering
\begin{tabular}{ c | c | c }

  {\bf Quantity} & {\bf Formula} & {\bf Value} \\
  \hline
  
  Coulomb logarithm (log$\Lambda$) &
  {\tiny $23.5 - \log n_{\text{e}}^{1/2}T_{\text{e}}^{-5/4}$ $ - \sqrt{10^{-5} + \frac{(\log T_{\text{e}} - 2)^2}{16}}$}
  & 7.2 \\
  
  Mass density ($\rho$)  &  
  $1.7 \times 10^{-24} (\sum_{j} M_jn_j)$ 
  & $1.6\times 10^{-3}$ g cm$^{-3}$ \\
  
  Debye length ($\lambda_{\text{D}}$) & 
  $ 7.4\times 10^2 \frac{T_{\text{e}}^{1/2}}{n_{\text{e}}^{1/2}} \big[ 1 + \frac{T_{\text{e}}}{T_{\text{i}}} Z_{\text{eff}} \big]^{-1/2} $ 
  &  $4.2 \times10^{-7}$ cm \\
  
  Sound speed ($c_{\text{s}}$) & 
  $9.8 \times 10^5 \frac{ [(\langle Z \rangle + 1) \gamma_{\text{I}} T_{\text{e}}]^{1/2}}{ \langle M \rangle ^{1/2}}$ 
  & $3.5 \times 10^7$ cm s$^{-1}$ \\
  
  Mach number & 
  ${\text{v}}_{\text{turb}}/c_{\text{s}}$ 
  &  0.6 \\
  
  Plasma $\beta$ & 
  $4.0 \times 10^{-11} \frac{n_{\text{e}}T_{\text{e}} + \sum_{j} n_j T_{\text{i}} }{ B_{\text{RMS}}^2}$
  & 44 \\
  
  H-ion mean free path ($\lambda_{\text{Hion}}$) & 
  $2.1 \times 10^{13} \frac{T_{\text{i}}^2}{ \big( Z_{\text{H}}^2Z_{\text{eff,B}}n_{\text{e}}\log \Lambda \big)} $
  & $1.3 \times 10^{-3}$ cm \\
  
 Electron-ion mean free path ($\lambda_{\text{e}}$) & 
 $2.1 \times 10^{13} \frac{T_{\text{e}}^2}{ \big( Z_{\text{eff}}n_{\text{e}}\log \Lambda \big)}$
 & $1.2 \times 10^{-3}$ cm \\
  
 Electron Larmor radius ($r_{\text{g}}$) & 
 $2.4 \frac{T_{\text{e}}^{1/2}}{ B_{\text{RMS}}}$
 & $1.0 \times 10^{-4}$ cm \\
 
 Hydrogen Larmor radius ($\rho_{\text{H}}$) & 
 $1.0 \times 10^2 \frac{M_{\text{H}}^{1/2}T_{\text{i}}^{1/2}}{Z_{\text{H}}B_{\text{RMS}}}$
 & $4.1 \times 10^{-3}$ cm \\
 
 Unmagnetized thermal diffusivity ($\chi_{\text{S}}$) & 
 $3.0\times 10^{21} \frac{T_{\text{e}}^{5/2}}{Z_{\text{eff}}n_{\text{e}}\log \Lambda}$
 & $5.9 \times 10^6$ cm$^2$ s$^{-1}$ \\
 
 Magnetized thermal diffusivity ($\chi_{\text{m}}$) & 
 $ {\chi_{\text{S}} r_{\text{g}}} / {\lambda_{\text{e}}}$
 & $4.7 \times 10^5$ cm$^2$ s$^{-1}$ \\
 
  Suppressed thermal diffusivity ($\chi$) & 
 $ {\sim} \chi_{\text{S}}/150 $
 & $3.9 \times 10^4$ cm$^2$ s$^{-1}$ \\
 
 Turbulent P\'eclet number (Pe$_{\text{turb}}$) & 
 ${\text{v}}_{\text{turb}}L / \chi$
 & ${\sim}30$ \\
 
 Dynamic viscosity ($\mu$) & 
 {\tiny $4.27\times 10^{-5} \frac{M_{\text{H}}^{1/2} T_{\text{i}}^{5/2} n_{\text{H}}}{\log \Lambda Z_{\text{eff,B}}n_{\text{e}}}$} 
 & 7 g cm$^{-1}$ s$^{-1}$ \\
 
 Kinematic viscosity ($\nu$) & 
 $\mu / \rho $ 
 & $4.3 \times 10^3$ cm$^2$ s$^{-1}$ \\
 
 Turbulent Reynolds number (Re$_{\text{turb}}$) & 
 ${\text{v}}_{\text{turb}}L/\nu$
 & 280 \\
 
 Viscous dissipation scale ($l_{\nu}$) & 
 $L/{\text{Re}}_{\text{turb}}^{3/4}$ 
 & $8.8 \times 10^{-4}$ cm \\
 
 In-flow Resistivity ($\eta_{||}$) & 
 $3.1\times 10^5\frac{Z_{\text{eff}} \log \Lambda}{T_{\text{e}}^{3/2}}$
 & 350 cm$^2$ s$^{-1}$ \\
 
 Magnetic Reynolds number (Rm$_{\text{turb}}$) & 
 ${\text{v}}_{\text{turb}}L/\eta_{||}$
 & $3.5 \times 10^3$ \\
 
 Magnetic Prandtl number (Pm) & 
 Rm / Re 
 & 12 \\
 
 Resistive dissipation scale ($l_{\eta}$) & 
 $L/{\text{Pm}}^{1/2}$
 & $2.5 \times 10^{-4}$ cm \\
  
\end{tabular}
\caption{Summary of derived plasma parameters at t $= 25$ ns after the start of the drive laser pulse. The values are reported in CGS.}
\label{table:values2}
\end{table}

\newpage

\section*{Gated X-ray detector to diagnose electron temperatures}

Here we introduce a passive diagnostic configured to measure spatially-resolved electron temperature profiles using a gated 
X-ray detector (GXD), which includes an X-ray framing camera, differential filtering, a pinhole array, and two sets of collimator arrays.

The GXD was mounted onto a diagnostic snout to diagnose a side profile of the plasma flows as illustrated in Figure 1 of the main text. 
The GXD was equipped with two sets of collimator arrays, both of which comprised of a staggered array of four collimators, 250 µm in diameter.  
A pinhole array comprising a staggered array of four pinholes, 25 $\mu$m in diameter, was positioned between the 
two sets of collimator arrays, at a distance that enabled a 2$\times$ magnification of the interaction region.  
Each of the four pinholes were concentrically aligned to each respective collimator of both collimator arrays and 
staggered to avoid overlap of images onto 4-strips of the X-ray framing camera.  
All four of the strips were independently timed to enable the temporal characterization of the plasma.  
At least two of the pinholes were equipped with distinct filters to uniquely attenuate the X-ray plasma 
emission associated with each pinhole at a similar time (e.g., t = 25 ns as illustrated in Figure 2 of the main text) before 
imaging onto the camera.  In the absence of an absolute calibration of the X-ray framing camera, a direct comparison of the uniquely-attenuated 
X-ray plasma emission from the two pinholes enabled a characterization of the electron temperature.  
The distinct filters were positioned on a kinematic base, in front of the camera.  For most shots, 6.56 $\mu$m of polyimide (Poly) 
covered half of the camera and two of the staggered pinholes, each of them independently timed 
with gate width of 100 ps. The second half of the camera was covered with 2.38 $\mu$m of vanadium (V) 
to filter the remaining two pinholes, which were co-timed with first two.
For each shot we therefore collected four X-ray images: two Poly-filtered images taken at two distinct times, 
and two V-filtered images co-timed with the Poly-filtered images. 
Based on the magnification of the GXD, the expected spatial resolution of the X-ray images is 50 $\mu$m.  We also note that for a turbulent velocity of $2\times10^7$ 
$\mathrm{cm}\, \mathrm{s}^{-1}$, the blurring induced by the turbulent motions within the gate width of the camera is 20 $\mu$m.

\begin{figure}[!h]
    \includegraphics[width=6.in]{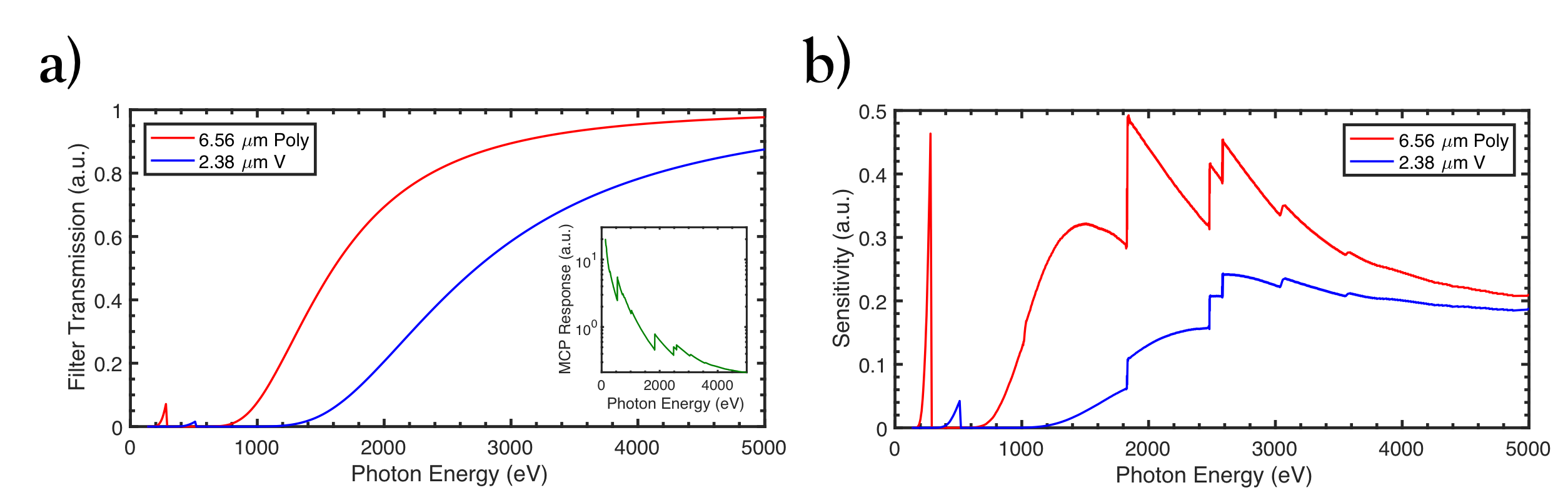}
\caption{{\bf (a)} Filter transmission curves, with reference to the Center for X-Ray Optics at Lawrence Berkeley National Laboratory database, for 6.56 $\mu$m thick polyimide (Poly) and 2.38 $\mu$m thick vanadium (V) used in experimental shots. The inset shows the relative microchannel plate (MCP) response of the NIF X-ray framing cameras \cite{rochau}.  {\bf (b)} Sensitivity curves for each filter, determined by comparing each transmission curve to the MCP response shown in (a).}
\label{sensitivity}
\end{figure}

The sensitivity curves of the GXD diagnostic take into account both the relative MCP response of the camera and the filter transmission of each filter chosen for a particular shot.  In Figure \ref{sensitivity}b, the sensitivity was calculated by multiplying the filter transmission with the MCP response for each discrete photon energy.  We specifically chose filters (i.e., Poly and V) that were both compatible with the facility requirements and had significantly different transmission curves.  

\begin{figure}[!h]
    \includegraphics[width=6.in]{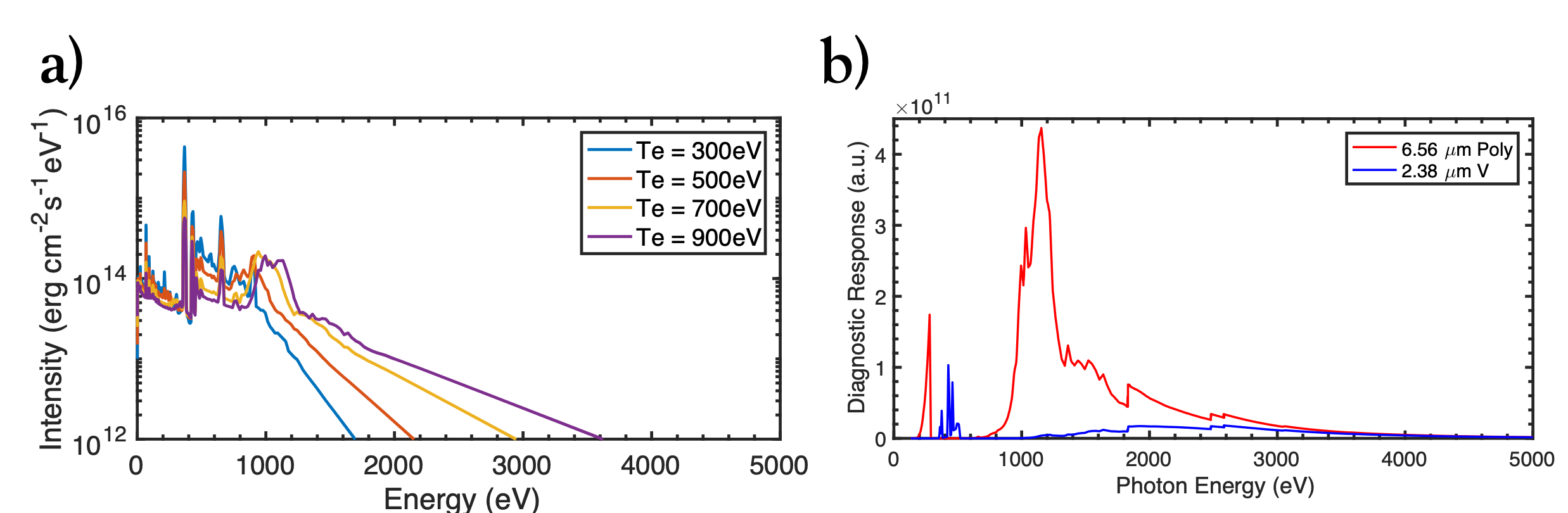}
\caption{{\bf (a)} SPECT3D calculated X-ray spectra, assuming an electron density of $10^{20}$ cm$^{-3}$.  {\bf (b)} Expected frequency-resolved signal measured by the GXD diagnostic for $T_e=900$ eV and $n_e=10^{20}$ cm$^{-3}$.}
\label{response}
\end{figure}

We modeled the X-ray spectrum emitted from a region of given density (between $10^{19}$ and $10^{21}$ cm$^{-3}$) and temperature (in the range 200--5000 eV) using SPECT3D (www.prism-cs.com).  SPECT3D is a collisional-radiative spectral analysis code designed to simulate the atomic and radiative properties of laboratory and astrophysical plasmas. 
Given the multi-component nature of the plasma, we used PrOpacEOS (Prism opacity and equation of state code) equation of state tables, with non-LTE atomic transitions to generate simulated X-ray spectra, an example of which is shown in Figure \ref{response}a.

\begin{figure}[!h]
\centering
    \includegraphics[width=4.in]{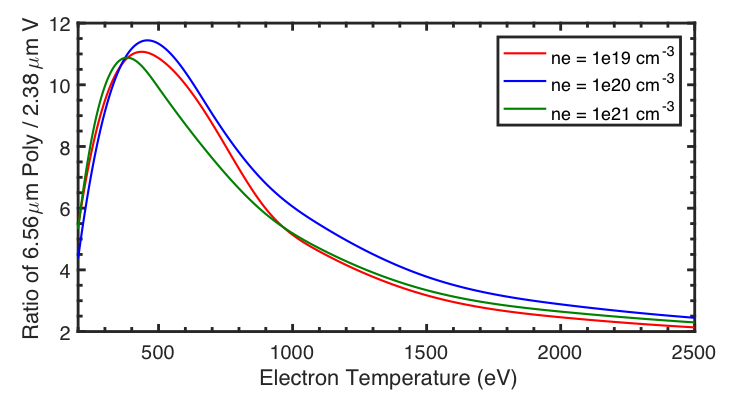}
\caption{The ratio curves of 6.56 $\mu$m polyimide and 2.38 $\mu$m vanadium are displayed for different electron densities.}
\label{ratiocurve1}
\end{figure}

For many different combinations of electron temperatures  and densities, the synthetic X-ray spectra 
were used to determine 
expected frequency-resolved signal of each filter.  In the example of Figure~\ref{response}b, an expected frequency-resolved signal for each filter was calculated by multiplying a simulated X-ray spectrum for 
$T_e=900$ eV and $n_e=10^{20}$ cm$^{-3}$
with each respective sensitivity (shown in Figure~\ref{sensitivity}b).  A ratio of the two signals was calculated and plotted as a discrete point of the curves in
Figure~\ref{ratiocurve1}. This process was repeated for numerous combinations of electron temperatures and densities to calculate the ratio curves in Figure~\ref{ratiocurve1}. The latter relate the electron temperature of the plasma with a ratio of the X-ray plasma emission collected by the GXD for each filter (i.e., a ratio of Poly-filtered emission to V-filtered emission), for a fixed density. 
Note that opacity effects can be neglected and the emission is optically thin for all plasma conditions considered here.

From Figure \ref{ratiocurve1}, we first note that the shape of the curves is rather insensitive to the electron density, and mostly depends on the electron temperature.
Moreover, the choice of filters results in a peak in the ratio curve near $T_e\sim$450 eV.  If we exclude temperatures below 450 eV on the basis that the plasma is expected to be much hotter (as indicated by FLASH simulations), then there is a one-to-one correspondence between the measured emission ratio and the plasma's electron temperature.
Therefore, in a plasma with uniform density and temperature, the value of the filter ratio can be use to infer an electron temperature. 
For $T_e > 2.5$ keV, the ratio curve becomes flat with temperature, thus determining the upper sensitivity bound.

\color{black}

For a plasma with spatially varying emission -- which itself arises directly from variations in temperature and density -- an interpretation of the two-dimensional temperature maps recovered using our technique is required. In the special case when the emission predominantly arises from a volume in \color{black} the \color{black} plasma with only small variations in density and temperature with respect to some mean value, a reasonable interpretation for the maps is that of the electron temperature, mass-averaged along the line of sight. On the other hand, if there are large departures from the mean value of temperature (and particularly if those values surpass the reported sensitivity bounds), this interpretation becomes less reliable. We \color{black} verify both of \color{black} these claims using the 3D FLASH simulations of our experiment described in the next section.

\color{black}

\begin{figure}[!h]
\centering
    \includegraphics[width=6.in]{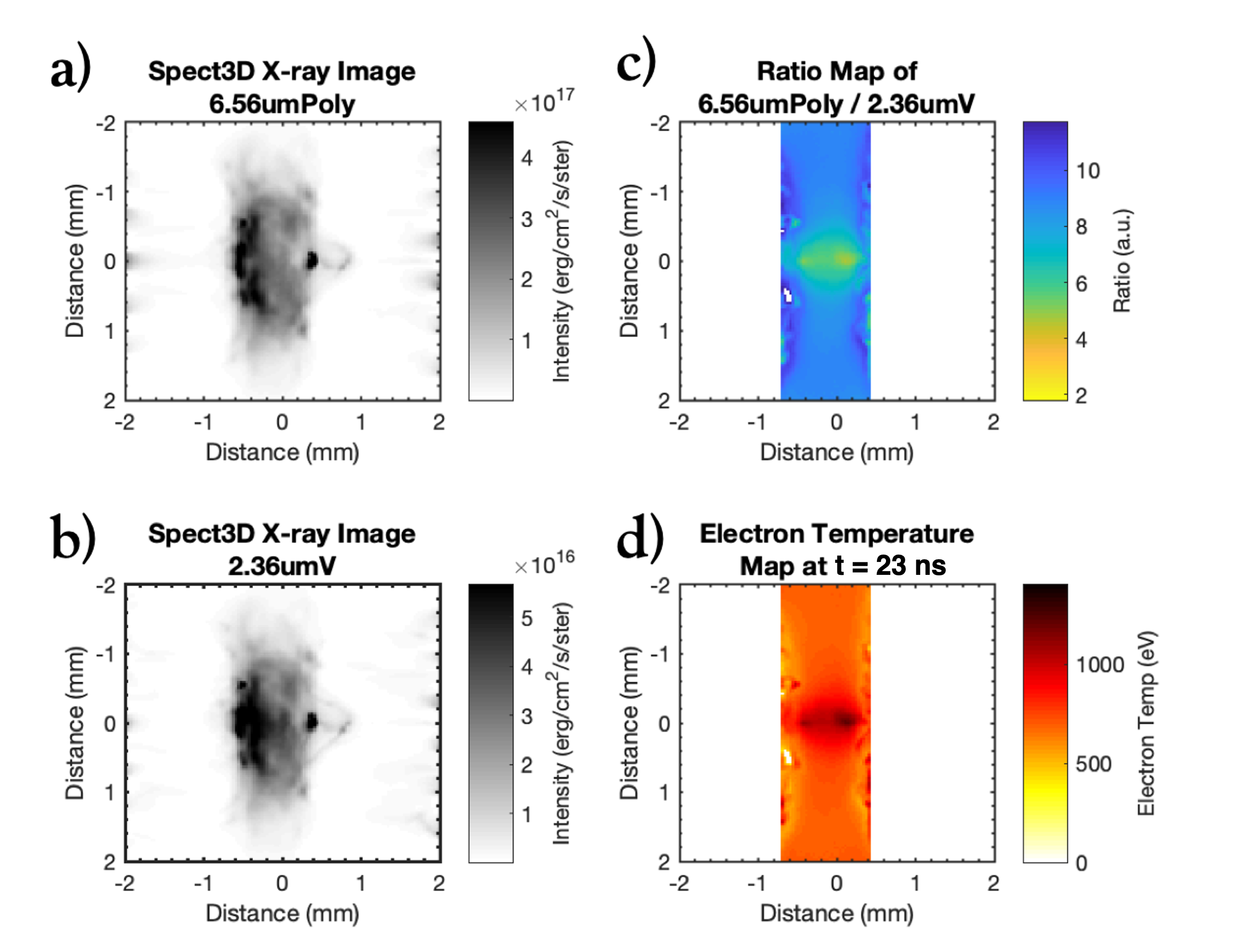}
\caption{ {\bf (a)} and {\bf (b)} show 2-D synthetic X-ray images from the conduction-on FLASH simulations, accounting for the two different filter transmissions.
The intensity ratio map is shown in {\bf (c)}. The ratio map has been calculated by ignoring data outside the central interaction region (that is, points with horizontal coordinate lower than -1 mm and larger than 1 mm have been ignored).  {\bf (d)} The electron temperature is extracted from the X-ray intensity ratio as shown in Figure~\ref{ratiocurve1}.}
\label{spect3d1}
\end{figure}

\newpage
\subsection*{\color{black} Interpretation \color{black}of temperature measurements}

\color{black} We used simulations carried out with the FLASH code 
to design, execute, and interpret the experiments discussed here. While the FLASH simulations are described in detail in a companion paper, here we discuss how they were used to interpret the temperature measurements derived from the GXD.
FLASH \cite{Fryxell2000} is a publicly available, parallel, multi-physics, adaptive mesh refinement (AMR), finite-volume Eulerian hydrodynamics and magneto-hydrodynamics (MHD) code developed at the Flash Center for Computational Science. FLASH scales well to over a 100,000 processors, and uses a variety of parallelization techniques including domain decomposition, mesh replication, and threading to optimally utilize hardware resources. FLASH is professionally managed software with version control, coding standards, extensive documentation, user support, and integration of code contributions from external users. The code is subject to daily, automated regression testing on a variety of platforms. 
Over the past nine years, extensive High Energy Density Physics (HEDP) and extended-MHD capabilities have been added in FLASH \cite{tzeferacos2015}
as part of the U.S. DOE NNSA-funded FLASH HEDP Initiative and through support by Los Alamos National Laboratory (LANL) and Lawrence Livermore National Laboratory (LLNL). These include multiple state-of-the art hydrodynamic and MHD
shock-capturing solvers; a generalized Ohm's law that incorporates extended MHD terms of the Braginskii formulation \cite{braginskii1965}; three-temperature extensions with anisotropic thermal conduction, heat exchange,
multigroup radiation diffusion, tabulated multi-material equations of state and opacities, laser energy deposition, and numerous simulated diagnostics \cite{tzeferacos2017}. The FLASH code and its capabilities have been validated through benchmarks and code-to-code comparisons \cite{fatenejad2013b, orban2013}, as well as through direct application to laboratory experiments \cite{meinecke2014, falk2014,yurchak2014,tzeferacos2015,Li2016,tzeferacos2017,tzeferacos2018,Rigby2018,Lu2019,Gao2019, White2019, Chen2020, bott2021time, bott2021inefficient}. The latter include experiments carried out at the Omega Laser Facility that employed the TDYNO (turbulent dynamo) platform used in this work \cite{tzeferacos2018, Chen2020, bott2021time}. For a discussion on the evolution of the platform, see Ref. \cite{Muller2017}.
\color{black}

\begin{figure}[!h]
\centering
    \includegraphics[width=6.in]{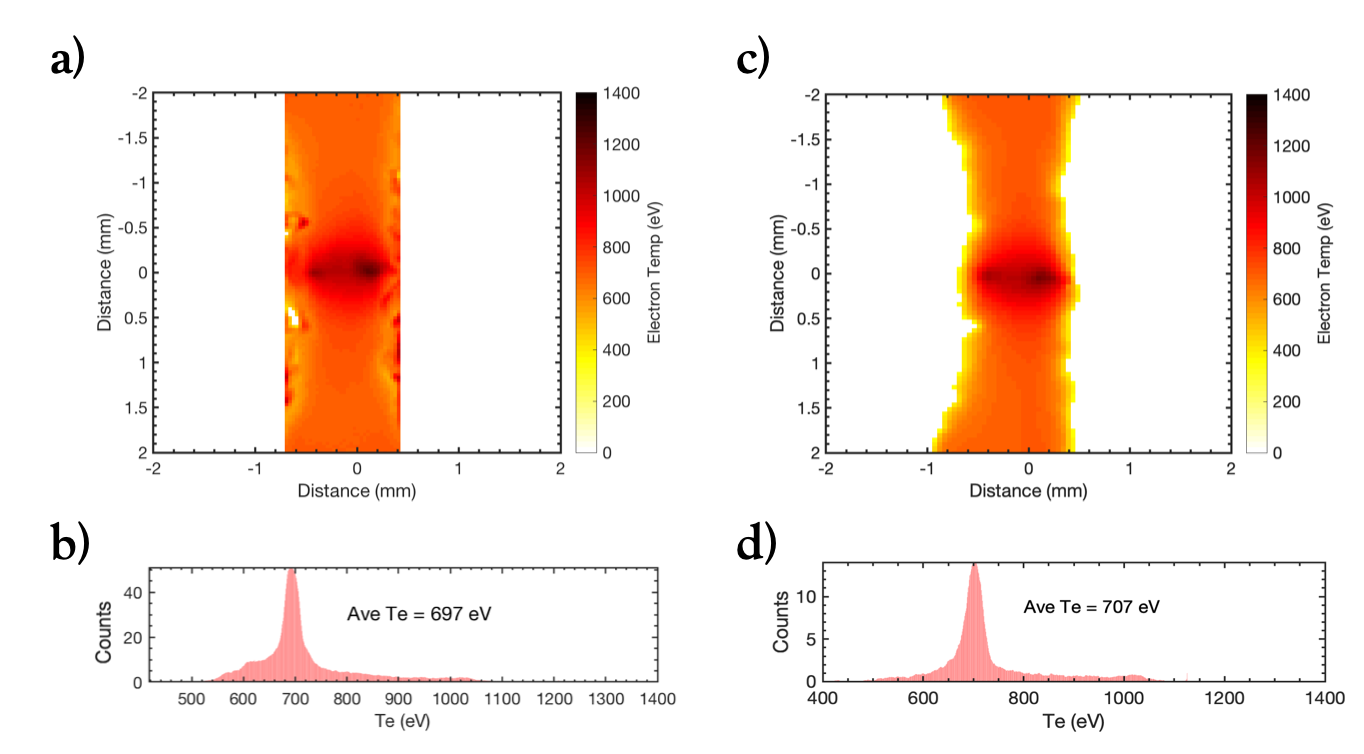}
\caption{{\bf (a)} Extracted temperature profiles from the synthetic X-ray images' intensity ratio, assuming a constant electron density $n_e=5.5\times 10^{20}$ cm$^{-3}$. {\bf (b)} Corresponding temperature distributions for panel {\bf (a)}. {\bf (c)} Mass-averaged temperatures, along the GXD detector line of sight, obtained from FLASH simulations at $t=25$ ns. {\bf (d)} Corresponding temperature distributions for panel {\bf (c)}. }
\label{spect3d2}
\end{figure}

{\bf Conduction-on case.} \color{black} To interpret the temperature measurements, we first consider \color{black} conduction-on FLASH simulations (i.e., with the inclusion of Spitzer conductivity in the system of equations evolved by FLASH \cite{tzeferacos2015}), which exhibit smooth interaction-region temperature profiles with small fluctuations with respect to the mean value. We post-processed the results of these simulations at $t=23$ ns with SPECT3D to produce a synthetic X-ray image of the interaction region as seen by the GXD diagnostics. SPECT3D was run with various filters in front of the detector. The resultant images are shown in Figure \ref{spect3d1}, with a $50$ $\mu$m spatial smoothing applied to mimic the instrument resolution.  As we did with the experimental images, we first create a map of X-ray intensity ratios, and then, taking $n_e=5.5\times 10^{20}$ cm$^{-3}$ (as measured at this time with optical Thomson scattering), construct an electron-temperature map.

The comparison between the inferred GXD temperature map from the synthetic X-ray images and the temperature map obtained by mass averaging the electron temperature (from FLASH simulations) along the same line of sight is shown in Figure \ref{spect3d2}.
The two images are, indeed, remarkably similar, giving us confidence in the diagnostic approach. We observe an average temperature of 648 eV and a peak temperature of $\sim 1$ keV with the synthetic GXD diagnostics. From the FLASH simulations, we obtain 
a mass-averaged temperature of 697 eV with a peak of $\sim 1.1$ keV, which is almost identical to the previous values.  Furthermore, the temperature distributions in the two images are  quantitatively very similar, as shown in Figures \ref{spect3d2}b and \ref{spect3d2}d.

\begin{figure}[!h]
\centering
    \includegraphics[width=6.in]{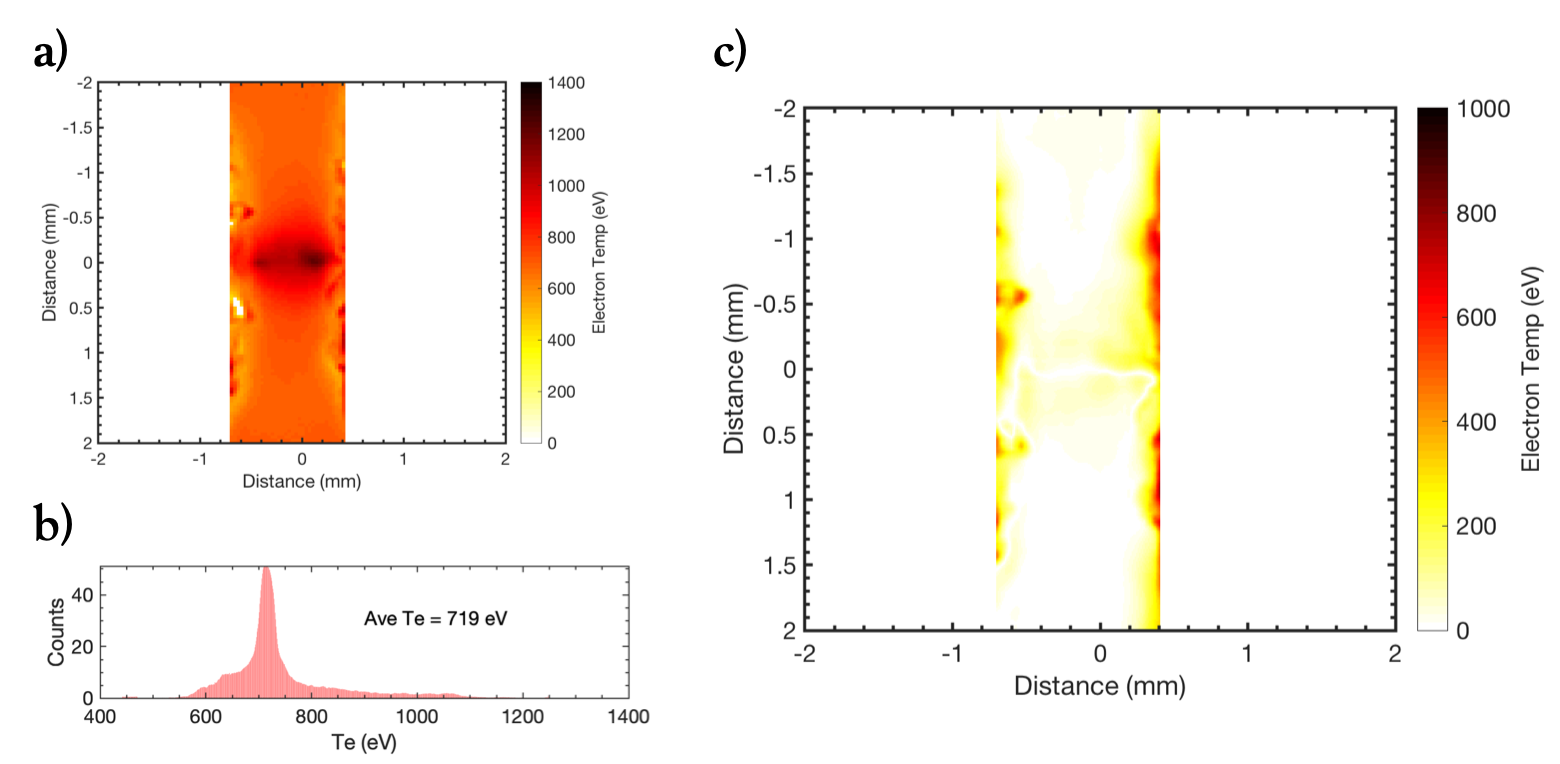}
\caption{{\bf (a)} Same as in Figure 5a but assuming $n_e=2.8\times 10^{20}$ cm$^{-3}$. {\bf (b)} Temperature distributions corresponding to the map shown in panel {\bf (a)}. {\bf (c)} Absolute temperature difference between panel {\bf (a)} and Figure \ref{spect3d2}a.}
\label{spect3d3}
\end{figure}

The differences between the GXD calculated temperature and the mass-averaged FLASH temperatures are mainly attributed on the assumption of constant density in the GXD retrieval, as illustrated in Figure \ref{ratiocurve1}.
On average, the estimated error associated with this assumption is $< 10\%$. 
This can be seen in Figure \ref{spect3d3}.
By assuming a lower value for the density, 
$n_e=2.8\times10^{20}$ cm$^{-3}$ (that is, 
50$\%$ lower than before), we can achieve an even more accurate electron-temperature map.  A comparison between this GXD predicted map and the FLASH temperatures produces an error in the central interaction region that is $< 5\%$ (see Figure \ref{spect3d3}c).  
 
\begin{figure}[!h]
\centering
    \includegraphics[width=6.in]{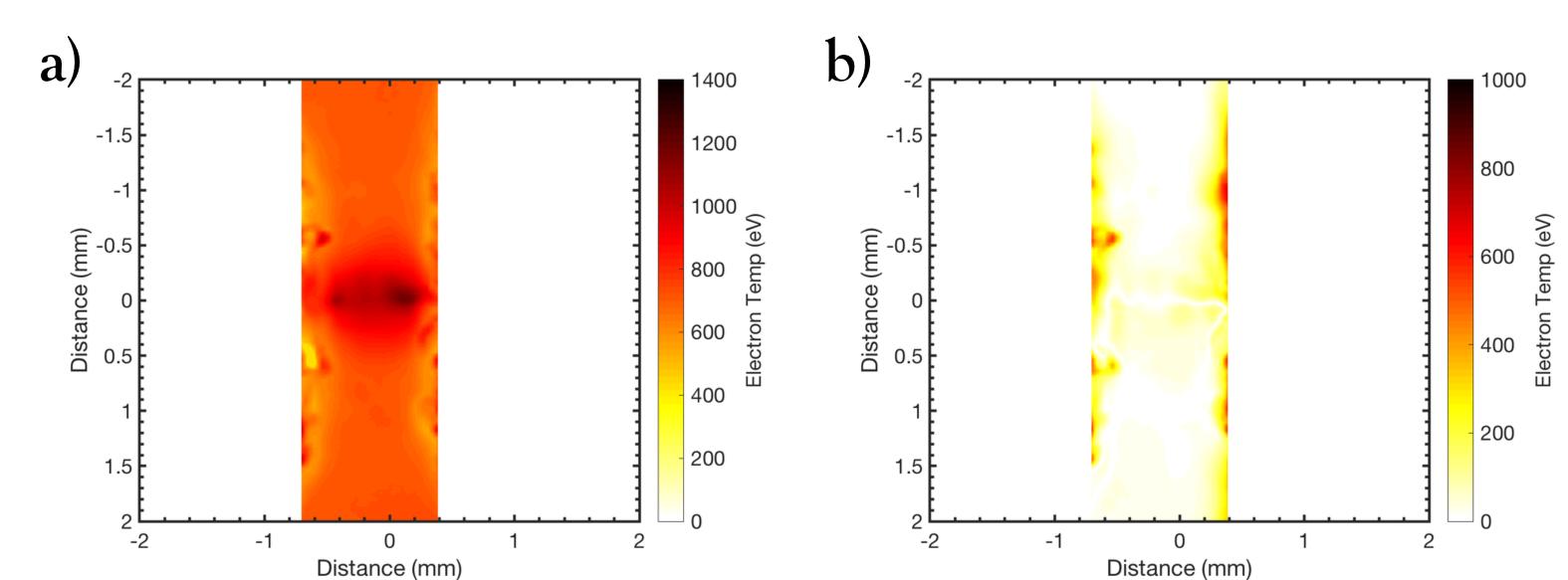}
\caption{{\bf (a)} Same as in Figure 5c but taking the FLASH averaged density along the line of sight at each position. {\bf (b)} Absolute temperature difference between panel {\bf (a)} and Figure \ref{spect3d2}a.}
\label{spect3d4}
\end{figure}

Finally, instead of using an average electron density for the whole region, we can assume a different (line-of-sight averaged) electron density for each position along the map. Taking these averaged densities from FLASH simulations and retrieving the electron temperature from the synthetic X-ray intensity ratios, we arrive at the map shown in Figure \ref{spect3d4}.
This shows marginal further improvement over the previous estimates, again reinforcing the point that the analysis developed here seems rather robust with respect to the particular value of electron density in the plasma.  

{\bf Error analysis.}
With our temperature diagnostics having been validated using synthetic data, we look now at the measured data and attempt to provide a full error analysis.  When dealing with the experimental data, we need to account for a few additional manipulations compared to what we did with the simulations.  First, to each measured X-ray image we  apply a background subtraction. Secondly, while the GXD snout had a nominal 25 $\mu$m diameter pinhole array, we have to account for possible variations (of order of $10\%-20\%$)
in their size.

\begin{figure}[!h]
\centering
    \includegraphics[width=5.8in]{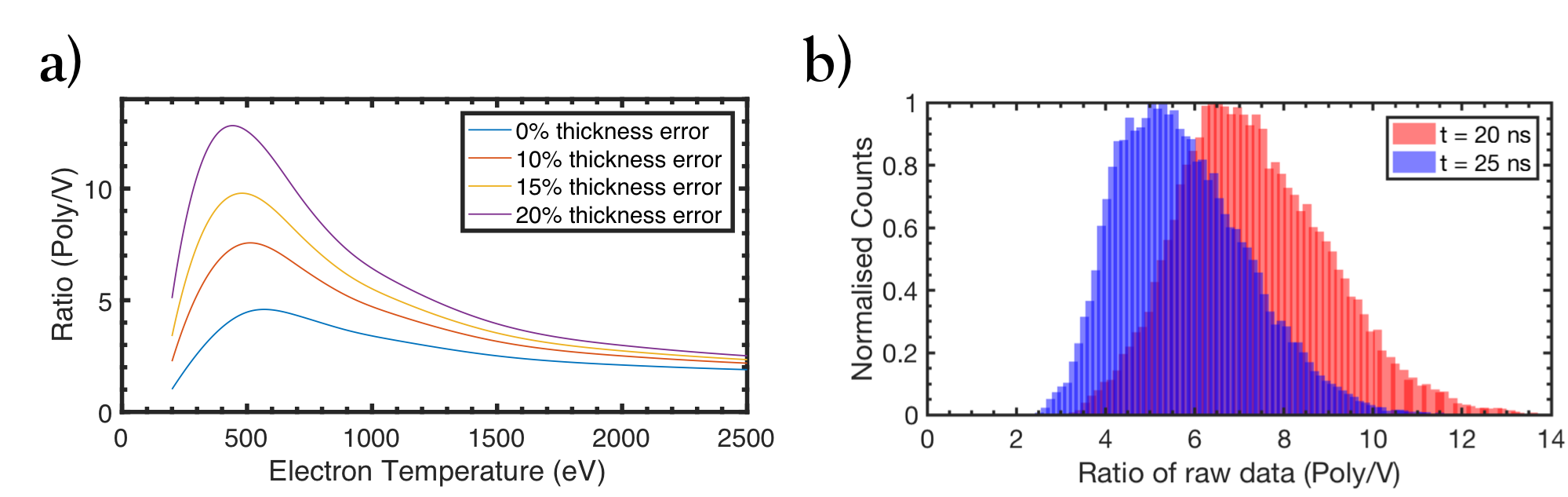}
\caption{{\bf (a)} Calculated intensity-ratio curves assuming errors in the nominal filter thickness (the filters' manufacturer, GoodFellow, advises that a $\pm 20 \%$ error in the supplied materials should be accounted for). {\bf (b)}  Intensity histograms for the NIF measurements at two different times.  We note ratio values ranging from $2.2 - 12$. Such a range can only be obtained assuming 18\% error in the filter thickness. This selects the ratio curve used in the data analysis.}
\label{gxdfig9}
\end{figure}

Errors in the filter thickness must also be taken into consideration (they can be as high as 20\%, according to the filters' manufacturer). 
In Figure \ref{gxdfig9}a we plot the X-ray
intensity ratio, assuming variations in the filters' thickness. We see clearly that the range of intensity ratios can vary depending on the exact thickness of the filters. 
While this would normally be a problem, we were able to deduce the actual filter thickness from the data itself.   
Plotted in Figure \ref{gxdfig9}b is the distribution of X-ray intensity ratios in the interaction region at two different times.  This plot shows a range of values from $2.2$ to $12$.  Assuming that the filter thickness does not  vary significantly in space, we then look at the available ratio curves and determine
which of the curves in Figure \ref{gxdfig9}a best reproduces this range.

In the main text, we compare temperature maps obtained on NIF to those recorded on the Omega laser facility. The GXD diagnostic is identical for the two experiments, with the only difference being that the Omega  filter combination was 2 $\mu$m mylar plus $40$ nm Al  and 4 $\mu$m mylar plus $80$ nm Al, and the ratio curve was obtained assuming the average electron density $n_e=10^{19}$ cm$^{-3}$.

\begin{figure}[!h]
\centering
    \includegraphics[width=6.in]{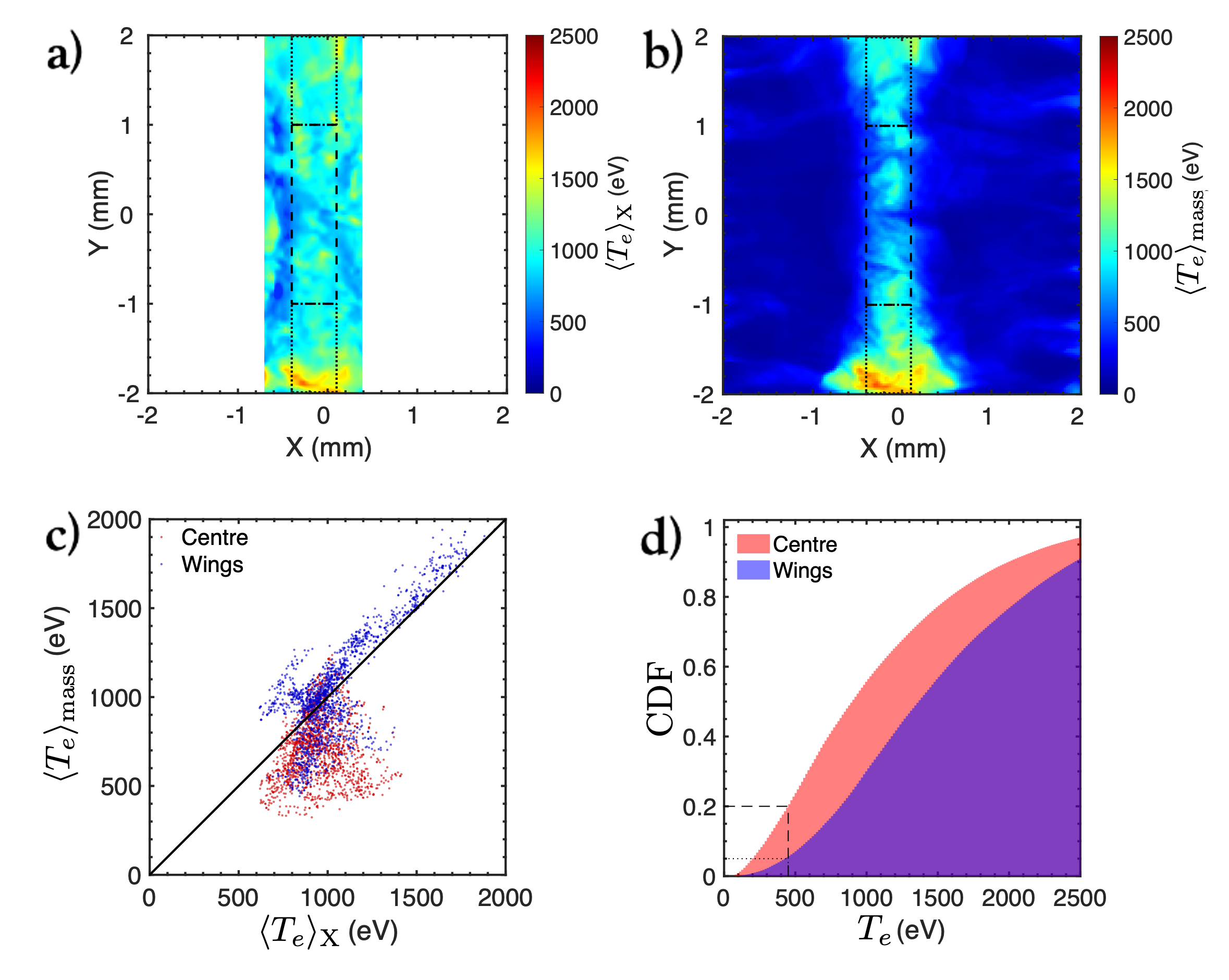}
\caption{{\bf (a)} 2D slice plot of X-ray brightness temperature $\langle T_e \rangle_{X}$ derived from post-processed conduction-off FLASH simulations at $t = 23$ ns. The cross-section of the $2\times 2\times 0.4$ mm${}^3$ `central' fiducial volume discussed in the main text is shown (dashed lines), along with the cross-section of two $2\times 1\times 0.4$ mm${}^3$ `wing' fiducial volumes (dotted lines). {\bf (b)} Same as (a), but for line-of-sight, mass-averaged temperature $\langle T_e \rangle_{\rm mass}$. {\bf (c)} Scatter plot of $\langle T_e \rangle_{X}$ against $\langle T_e \rangle_{\rm mass}$ for all points in central and wing fiducial volumes. {\bf (d)} Cumulative density function of the electron temperatures for the central and wing fiducial volumes. The dashed and dotted lines mark the fraction of the central and wing fiducial volumes, respectively, which are above the ${\sim}$450 eV peak of the temperature-ratio curve.}
\label{spect3d5}
\end{figure}

{\bf Conduction-off case.} Next, we perform the same analysis on the conduction-off FLASH simulations (i.e., with the omission of Spitzer conductivity, switched off immediately prior to the formation of the turbulent interaction region) at $t = 23$ ns: we produce 2-D synthetic X-ray images using SPECT3D with the appropriate filters, take the ratio, and infer a map of X-ray brightness temperature $\langle T_e \rangle_{X}$ (see Figure \ref{spect3d5}a). 
This can again be compared with the electron temperature mass-averaged along the line of sight, $\langle T_e \rangle_{\rm mass}$, determined directly from the simulations (Figure \ref{spect3d5}b). Qualitatively, the maps remain similar, but quantitatively there is some discrepancy, particularly in the area corresponding to the central fiducial volume discussed in the main text; quantitative agreement is better in the wings of the interaction-region plasma. These findings can be illustrated using a pointwise scatter plot that compares $\langle T_e \rangle_{X}$ with $\langle T_e \rangle_{\rm mass}$ (Figure \ref{spect3d5}c) for both the center of the interaction-region plasma, and its wings. While the scatter points in the wings cluster close to the line $\langle T_e \rangle_{\rm mass}\approx \langle T_e \rangle_{X}$, the clustering for points in the central volume is less strong. The most probable explanation for the discrepancy stems from the presence of large temperature \color{black} fluctuations (and therefore density fluctuations, since the central region is in approximate pressure balance, as shown in the next subsection) \color{black} in the conduction-off FLASH simulations. For the central region, which has a slightly lower mean temperature than the wings, ${\sim}20\%$ of the volume has a temperature below the lower sensitivity bound of the diagnostic (${\sim} 450$ eV) on account of these fluctuations (Figure \ref{spect3d5}d), whereas the equivalent figure for the wings is only ${\sim}5\%$.  

We note that the electron temperature in the conduction-off FLASH simulations will be subject to a small amount of numerical diffusion, despite the absence of physical thermal conductivity. As a result, the effective P\'eclet number will not be infinite in the FLASH simulations. The latter were executed using a variant of the Piecewise Parabolic Method (PPM, \cite{COLELLA1984}), with an explicit diffusive flux added to the numerical fluxes. The diffusion coefficient associated with the energy flux (viz. equation 4.2 in Ref. \cite{COLELLA1984}) can therefore be used to provide an order-of-magnitude estimate of the effective numerical thermal diffusivity $\chi_{\text{n}}\propto K \nabla \cdot \textbf{v} \Delta^2$. Here $K$ is a non-dimensional constant ($K=0.1$ in the simulations), $\nabla \cdot \textbf{v}$ is the multidimensional divergence of the velocity (viz. equation 4.5 in Ref. \cite{COLELLA1984} and $\langle \nabla \cdot \textbf{v} \rangle \sim 4\times10^7\, \text{s}^{-1}$ in the interaction region), and $\Delta$ is the resolution element of the simulation ($\Delta \sim 25 \times 10^{-4}$ cm). This yields $\chi_{\text{n}} \sim 25\, \text{cm}^2\, \text{s}^{-1} \ll \chi$, and a numerical P\'eclet number $\text{Pe}_{\text{n}} = \text{v}_{\text{turb}}L/\chi_{\text{n}} = 4,800 \gg \text{Pe}_{\text{turb}}$.

\subsection*{Pressure balance and radiative cooling in the conduction-off FLASH simulations}

\begin{figure}[!h]
\centering
    \includegraphics[width=4.5in]{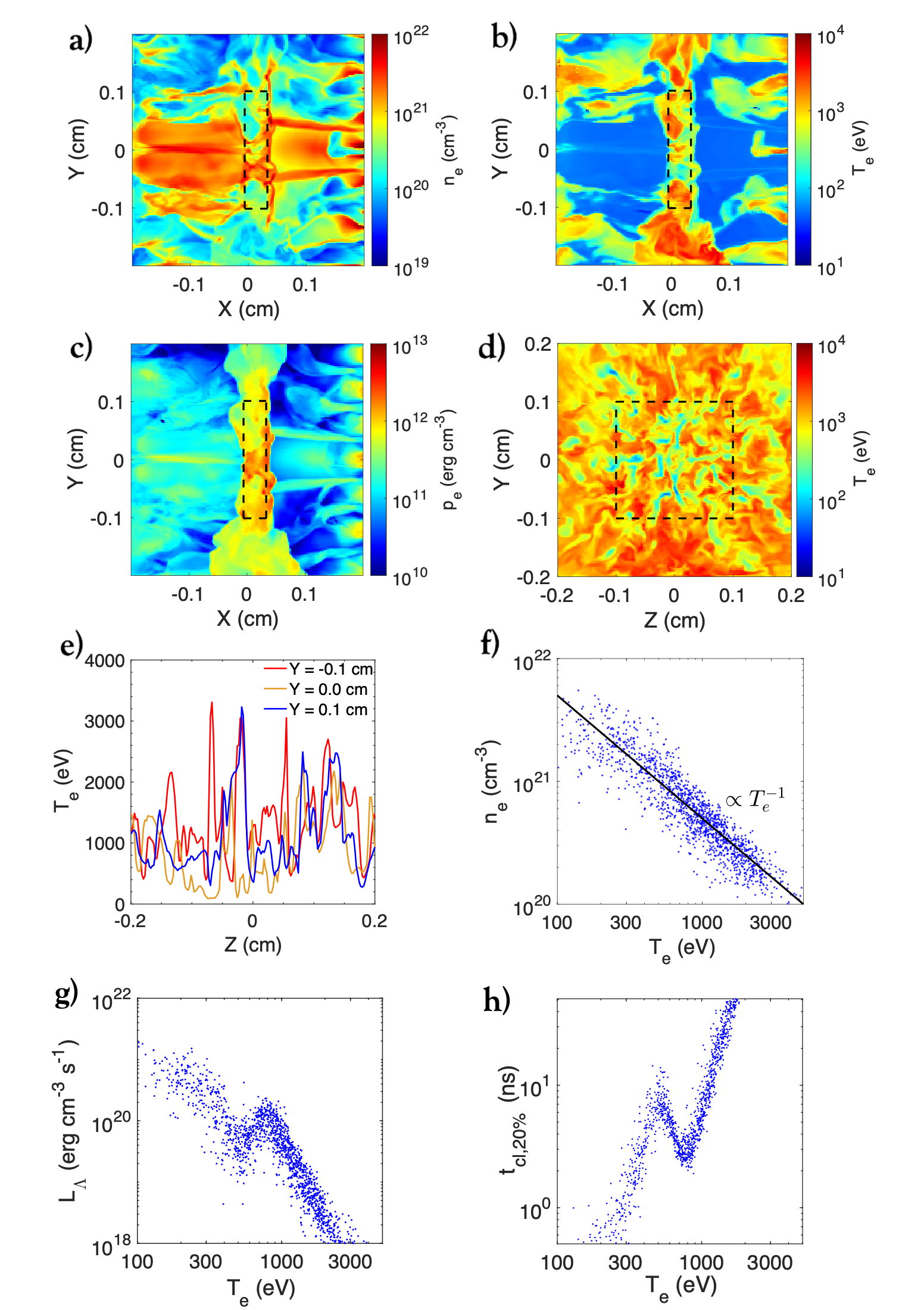}
\caption{{\bf (a)} 2D slice plot in the $Z = 0$ plane of the electron number density in the conduction-off FLASH simulation at $t = 23$ ns. The cross-sectional area of the $2\times 2\times 0.4$ mm${}^3$ `central' fiducial volume discussed in the main text is shown. {\bf (b)} Same as (a), but for electron temperature. {\bf (c)} Same as (a), but for electron pressure.
{\bf (d)} 2D slice plot in the $X = 0$ plane of electron temperature in the conduction-off FLASH simulation at $t = 23$ ns. The cross-sectional area of the $2\times 2\times 0.4$ mm${}^3$ `central' fiducial volume discussed in the main text is shown. {\bf (e)} Temperature profiles (in the $Z$ direction) in the $X = 0$ plane of electron temperature in the conduction-off FLASH simulation at $t = 23$ ns.   
{\bf (f)} Scatter plot of the electron temperature against the electron number density at 1,500 random selected points from the conduction-off FLASH simulations in the central fiducial volume. {\bf (g)} Same as f), but with scatter plot of the cooling function against electron temperature. {\bf h)} Same as f), but with scatter plot of the estimated time taken for the plasma to cool 20\% against electron temperature.}
\label{FLASHpress}
\end{figure}

In the main text, we claim that the central part of the interaction-region plasma is in approximate pressure balance; here, we verify that this is indeed the case in the conduction-off FLASH simulations of our experiment. Figures \ref{FLASHpress}a-c show 2-D plots of the electron number density, temperature and pressure in these simulations in the $X$-$Y$ plane.  
It can be seen by eye that, in spite of all quantities varying across the central part of the interaction-region plasma, fluctuations in density and temperature have a significantly larger magnitude than the pressure \color{black} (the large magnitude of electron temperature fluctuations in the interaction-region plasma is particularly evident in both 2D plots of this quantity in the $Y$-$Z$ plane (Figure \ref{FLASHpress}d), or profiles in the $Z$ direction at fixed values of $Y$ (Figure \ref{FLASHpress}e)). \color{black} More quantitatively, Figure \ref{FLASHpress}f shows a pointwise scatter plot of electron number density and temperature. We see that the latter is inversely correlated with the former over an (order-of-magnitude) range of values, and is consistent with the ideal electron pressure $p_e$ satisfying $p_e \propto n_e T_e \sim \mathrm{const.}$, as claimed.

We also claim in the main text that radiative-cooling instabilities play a role in generating and sustaining the temperature fluctuations in the interaction-region plasma. We support this claim here by calculating the cooling function $L_{\Lambda}$ in the conduction-off FLASH simulations, and then computing a pointwise scatter plot of $L_{\Lambda}$ against electron temperature in the simulated interaction-region plasma (Figure \ref{FLASHpress}g). The cooling function is indeed monotonically decreasing across much of the range of temperatures attained in the simulations, which is a necessary condition of radiative-cooling instabilities. We also calculate the cooling time at the same set of points in the simulation (Figure \ref{FLASHpress}h); we find that in at least some regions in the plasma, significant cooling occurs on a time scales that is comparable to dynamical time scales (${\sim}3$ ns), which supports our claim.       

\color{black}

\section*{Time-resolved Optical Thomson Scattering}
Optical Thomson scattering (OTS) was used to measure the electron plasma density at TCC, where the plasma flows collided.  For most shots, a frequency-tripled probe beam (351 nm light) was focused at TCC with a spot diameter of 1.2 mm and 12.8 kJ of energy delivered in a 8 ns square pulse.  The scattered light was collected in a cylindrical collection volume, 1.2 mm long and 50 $\mu$m diameter, with a scattering angle of 40$^{\text{o}}$.  A 1,200 grooves/mm grating spectrometer was used with a spectral bandwidth of approximately 64 nm, resulting in a dispersion of 4.27 nm/mm.  Calibration of the diagnostic was carried out using the 435.84 nm line from a HgAr lamp, fitted with a Gaussian instrument function of 0.291 nm.  

\begin{figure}[!h]
\centering
    \includegraphics[width=5.8in]{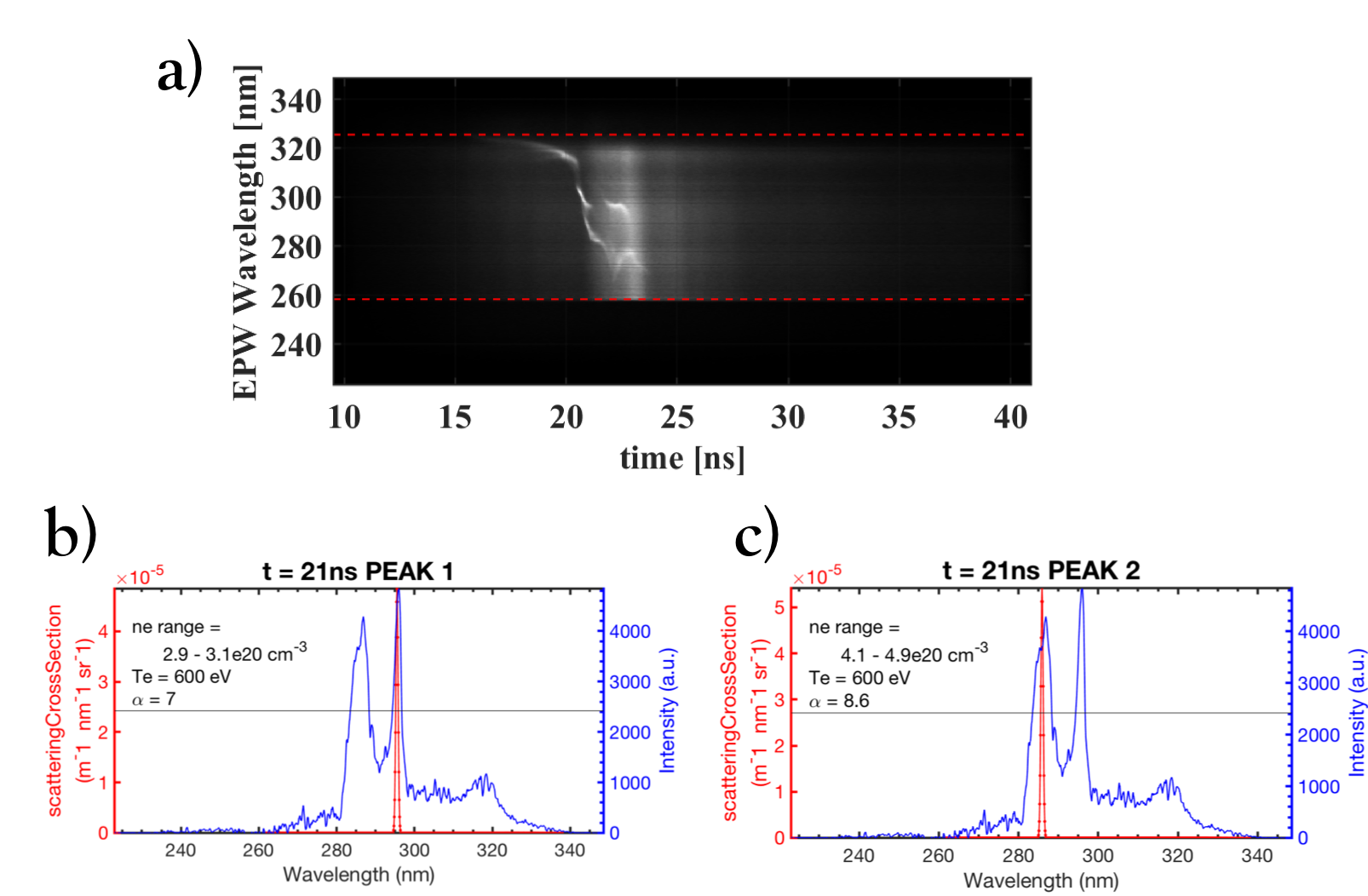}
\caption{{\bf (a)} Streaked Thomson scattering EPW data for the flow collision.  {\bf (b) and (c)} Lineouts at $t=21$ ns after the start of the drive beams. The lineouts are 10 pixel wide (226 ps). Superimposed on the data are the calculated EPW signals. Here, $\alpha = 1/k\lambda_{\text{De}}$, where $k$ is the scattering wavenumber and $\lambda_{\text{De}}$ the Debye length.
}
\label{ots1}
\end{figure}

Due to the highly turbulent plasma flow and large collection volume, we observe broad electron-plasma-wave (EPW) peaks and often multiple distinct peaks.  In a uniform plasma, the EPW peak width would be dominated by Landau damping, and therefore sensitive to the electron temperature.  In our experiment, however, the EPW peak width was influenced by the highly-turbulent plasma flow, and a value of the electron temperature could not reliably be determined from the OTS diagnostic. 
An example of the OTS data is shown in Figure \ref{ots1}. 
We see several distinct peaks at 21 ns after the start of the drive beams. 
The data therefore suggests that within the large collection volume, there are regions with electron density $\sim 3\times10^{20}$ cm$^{-3}$ and $\sim 5\times10^{20}$ cm$^{-3}$ surrounded by a broader distribution of densities.

\section*{Proton-imaging diagnostic to measure magnetic fields}

In this section, we provide further details of the analysis that we performed 
to measure both the root-mean-square (RMS) and maximum magnetic-field strengths attained in our experiment, as well as the field's correlation length $\ell_B$.

\begin{figure}[!h]
\centering
    \includegraphics[width=5.8in]{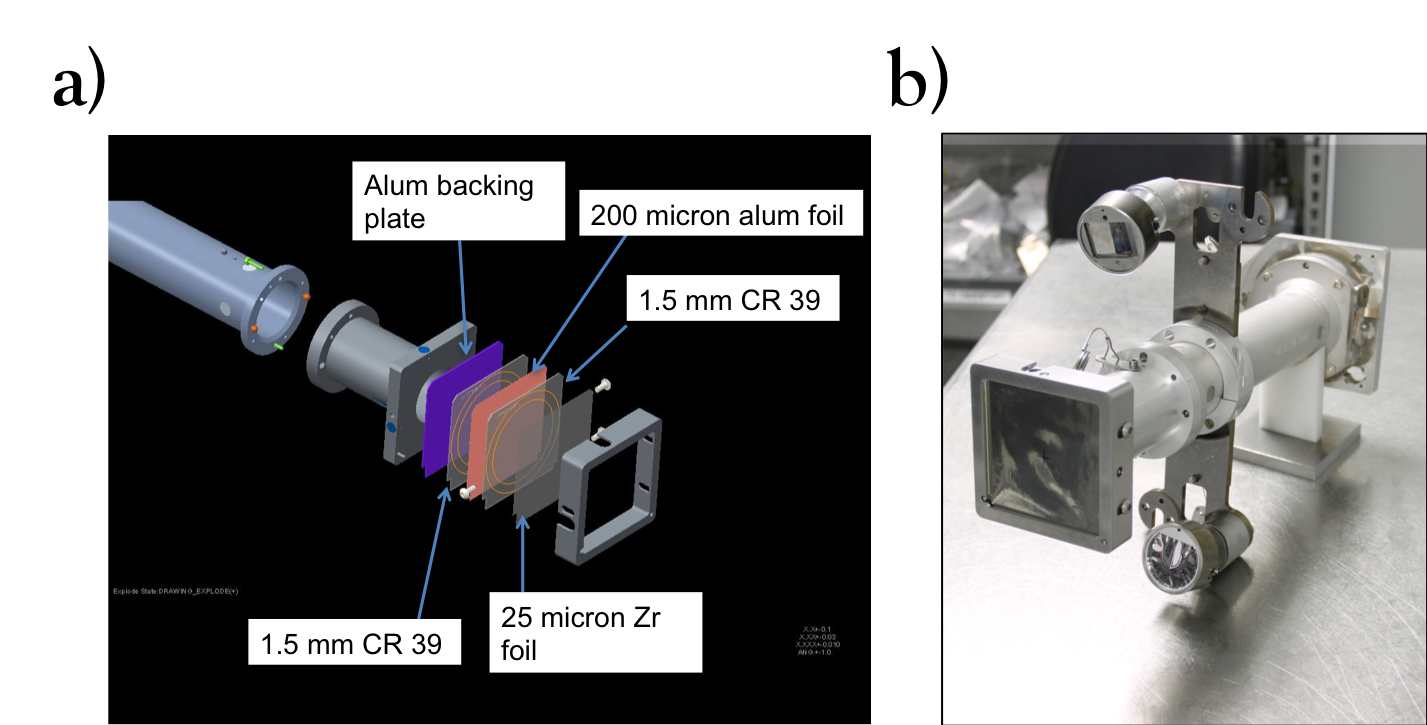}
\caption{{\bf Proton radiography (PRAD) diagnostic. (a)} Diagram of the filter arrangement. {\bf (b)} Photograph of the PRAD diagnostic snout.}
\label{prad1}
\end{figure}

The proton-radiography diagnostic uses a monoenergetic source of 14.7 MeV protons to backlight the plasma collision region, known to have high self-generated magnetic fields.  Every proton incident upon the 10 by 10 cm CR39 detector is observed, to  determine the distribution of electromagnetic fields present in the plasma. The filter pack shown in Figure \ref{prad1} is composed of two 1.5 mm thick CR-39 plates to measure both the 3 MeV and 14.7 MeV protons. To achieve the proper filtering, a $25 \pm 5$ $\mu$m Zr was placed on the front of the pack while an Al filter was placed between both CR-39 plates.  The thickness of each CR-39 plate can vary from 1.4-1.6 mm so the Al filter thickness was adjusted between $120-230 \pm 10$ $\mu$m to allow for 14.7 MeV protons to stop on the second CR-39 plate.  The overall energy uncertainty of the system is $\sim 0.34$ MeV.

The proton backlighter source is a 860 $\mu$m diameter capsule, consisting of a 3.1 $\pm$ 1 $\mu$m thick SiO$_2$ shell filled with D$_2$ gas at 6 atm and $^3$He at 12 atm, is located 18 mm from the midpoint of the two disks and ablated with 60 frequency-tripled laser beams with a 200 $\mu$m spot that deliver 43 $\pm$ 5 kJ in a 900 ps square pulse length.  Previous experiments indicate the importance of driving the capsule  symmetrically with high-energy beams to improve proton yield.  On average, the capsule was ablated with $15-20$ kJ from top beams and $  20-25$ kJ from bottom beams to allow for symmetry.  

\subsection*{Analysis of experimental data} \label{ExpAnalysis}

\begin{figure}[]
\centering
    \includegraphics[width=0.95\linewidth]{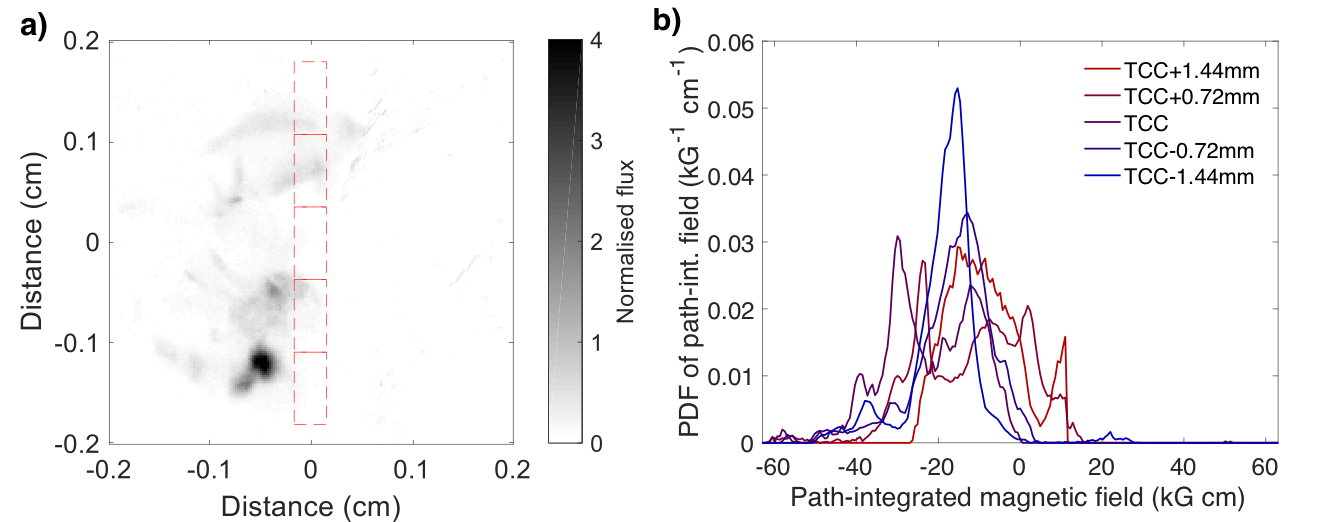}
\caption{\textbf{a)} 15.0 MeV proton image recorded at $t=25$ ns after the start of the laser drive, rotated so that the direction parallel to the slit is vertical. The normalization of the proton image is calculated relative to the mean flux of protons passing through the slit aperture that would have been measured in the absence of any magnetic fields. The imaging set-up has a $\times19$ magnification; for clarity's sake, we remove this factor, meaning that all lengths can be compared directly to the plasma's scale. \textbf{b)} Predicted distribution of the axial path-integrated fields at five distinct spatial locations along the slit. Here, the target chamber center (TCC) coincides with the geometric center of our target, and so with the center of the interaction region.}
\label{PRADfigSIA}
\end{figure}

As discussed in the Materials and Methods, the measurement of the magnetic field was performed by placing a 324 $\mu$m wide slit between the capsule and the colliding plasma region. Figure \ref{PRADfigSIA}a shows the outline of the slit onto the CR-39 plate.
Protons crossing the interaction region undisturbed (i.e., in the absence of any plasma or fields)  should end up on the CR-39 plate only within the slit outline.
By analyzing the deviations of the proton paths from their undisturbed trajectories, we can determine the mean, root-mean-square (RMS), and maximum value of the component of the path-integrated magnetic field that is co-aligned with the slit orientation (the length of the slit, or the `axial' component). Below, we explain our approach in greater detail.

Our approach is based on the fact that in a point-projection proton-imaging set-up such as that employed on our experiment, it can be shown (in the small-deflections limit) that the position $\boldsymbol{x}_{\perp s}$ of a proton arriving on the detector is related to its initial position $\boldsymbol{x}_{\perp 0}$ prior to passing through the plasma's magnetic field $\boldsymbol{B}$ via
\begin{equation}
\boldsymbol{x}_{\perp s} = \mathcal{M} \boldsymbol{x}_{\perp0} + \frac{e r_s}{m_p c V_0} \hat{\boldsymbol{n}} \times \int_0^{\ell_n} \mathrm{d} s \, \boldsymbol{B}\!\left[\boldsymbol{x}_0\!\left(s\right)\right] \, , \label{PRADSI_protproj}
\end{equation}
where $\mathcal{M}$ is the magnification of the imaging set-up, $e$ the elementary charge, $r_s$ the distance from the plasma to the detector, $m_p$ the proton mass, $c$ the speed of light, $V_0$ the initial proton velocity, $\hat{\boldsymbol{n}}$ the initial direction of the proton beam, $\ell_n$ the plasma's scale length, $s$ the path length of the proton through the plasma, and $\boldsymbol{x}_0\!\left(s\right)$ the trajectory of a proton with position $\boldsymbol{x}_{\perp 0}$ through the plasma in the absence of any fields. It follows that if the projected position $\mathcal{M} \boldsymbol{x}_{\perp0}$ of a given proton in the absence of magnetic fields is known, two components of the path-integrated magnetic field can be calculated by rearranging (\ref{PRADSI_protproj}):  
\begin{equation}
\int_0^{\ell_n} \mathrm{d} s \, \boldsymbol{B}_{\perp}\!\left[\boldsymbol{x}_0\!\left(s\right)\right] = -\frac{m_p c V_0}{e r_s} \hat{\boldsymbol{n}} \times \left[\boldsymbol{x}_{\perp s} - \mathcal{M} \boldsymbol{x}_{\perp0} \right]  \, , \label{PRADSI_protproj_rev}
\end{equation}
where $\boldsymbol{B}_{\perp} \equiv \boldsymbol{B} - (\boldsymbol{B}_{\perp} \cdot  \hat{\boldsymbol{n}})  \hat{\boldsymbol{n}}$ are the components of $\boldsymbol{B}$ perpendicular to $\hat{\boldsymbol{n}}$. A corollary of this result is that if the initial proton beam has a finite extent, the resulting distribution of protons is related directly to the distribution of (two components of) the path-integrated magnetic field. 

We apply this result to the slit-aperture proton-imaging data shown in Figure \ref{PRADfigSIA}a. 
The initial position with respect to the direction perpendicular to the slit orientation of imaging protons is
well constrained (i.e., on the central projected axis of the slit, with a $\pm171 \, \mu \mathrm{m}$ uncertainty arising from the slit's finite width), because any protons on the detector in the central location must have passed through the slit aperture prior to entering the plasma. Lineouts of the proton-flux distribution in the direction perpendicular to the slit orientation are therefore related to the distribution of the axial component of the magnetic field experienced by protons passing through the slit. The initial and final position with respect to the direction parallel to the slit orientation of imaging protons are not as strongly constrained, leading to an uncertainty in the position along the slit of where deflections have been acquired; we revisit the implications of this uncertainty later in this subsection.

\begin{figure}
\centering
    \includegraphics[width=0.7\linewidth]{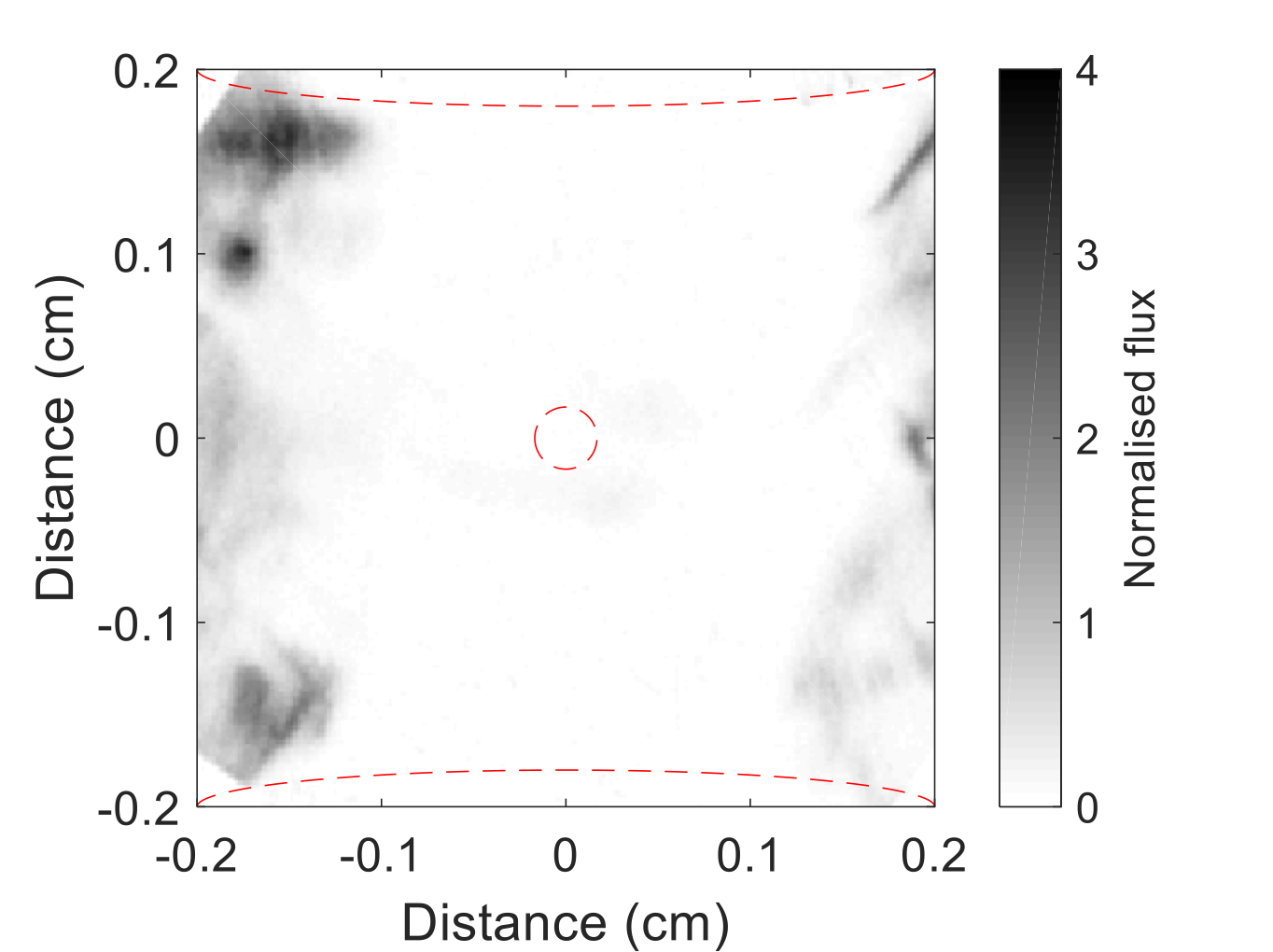}
\caption{15.0 MeV proton image of the colliding  plasma jets, but with all protons except those passing through a 300 $\mu$m pinhole blocked by a Cu shield. The outline of the pinhole (as well as the projected positions of the grids) are depicted by red dashed lines. The normalization of the proton image is calculated using the mean flux of protons passing through the pinhole that would have been measured in the absence of any magnetic fields. The fusion capsule that generated the proton beam in this case was imploded 25 ns after the initiation of the main laser drive.}
\label{PRADfigSIC}
\end{figure}

To elucidate further the structure of the magnetic field contained in the plasma, we divide the projection of the slit aperture into five equally sized regions, with the central position of a given region separated from adjacent regions by a distance $\ell_{\rm reg} = 0.72 \, \mathrm{cm}$ (see Figure \ref{PRADfigSIA}a).
We then calculate the one-dimensional distributions of protons in the direction perpendicular to the slit orientation, averaged over the region's other dimension. The resulting distributions can be converted to distributions of the axial path-integrated magnetic field in those five distinct regions. More specifically, using equation (\ref{PRADSI_protproj_rev}) and the known parameters of our imaging set-up ($V_0 = 5.3 \times 10^{9} \, \mathrm{cm/s}$, $\mathcal{M}\approx 19$, and $r_s = 28.2\, \mathrm{cm}$), a shift in proton position $\Delta \tilde{x} = |\boldsymbol{x}_{\perp s}/\mathcal{M} -  \boldsymbol{x}_{\perp0}|$ from its initial position (disregarding the magnification) is related to the path-integrated axial magnetic field via
\begin{equation}
\int_0^{\ell_n} \mathrm{d} s \, B_{\|} = 22 \left(\frac{\mathcal{M}}{19}\right) \left(\frac{V_0}{5.3 \times 10^{9} \, \mathrm{cm/s}}\right) \left(\frac{r_s}{28.2 \, \mathrm{cm}}\right)^{-1} \left(\frac{\Delta \tilde{x}}{0.6 \, \mathrm{cm}}\right)\mathrm{kG} \, \mathrm{cm}. \label{PRADSI_protproj_rev_nums}   
\end{equation}
The distributions of the axial path-integrated fields recovered for the five regions are shown in Figure \ref{PRADfigSIA}b. Finally, the mean, RMS, and maximum value of the axial component of the path-integrated magnetic field are calculated directly from its distribution function. 

We now reconsider the consequences of the uncertainty in the initial position along the slit of any given imaging proton. Both additional experimental data (see below) and the FLASH simulations of our experiment suggest that the magnetic fields in the interaction region are statistically isotropic. We therefore assume that the uncertainty in position is comparable to the RMS magnitude of $\Delta \tilde{x}$. This in turn can be used to provide a lower bound on the minimum separation $\ell_{\rm reg}$ that can reasonably be employed when determining the path-integrated axial magnetic field at spatial locations along the slit: $\ell_{\rm reg} \geq 0.6 \, \mathrm{cm}$. This is indeed smaller than our chosen distance between the five adjacent regions selected along the slit.  

\begin{figure}[!h]
\centering
    \includegraphics[width=0.95\linewidth]{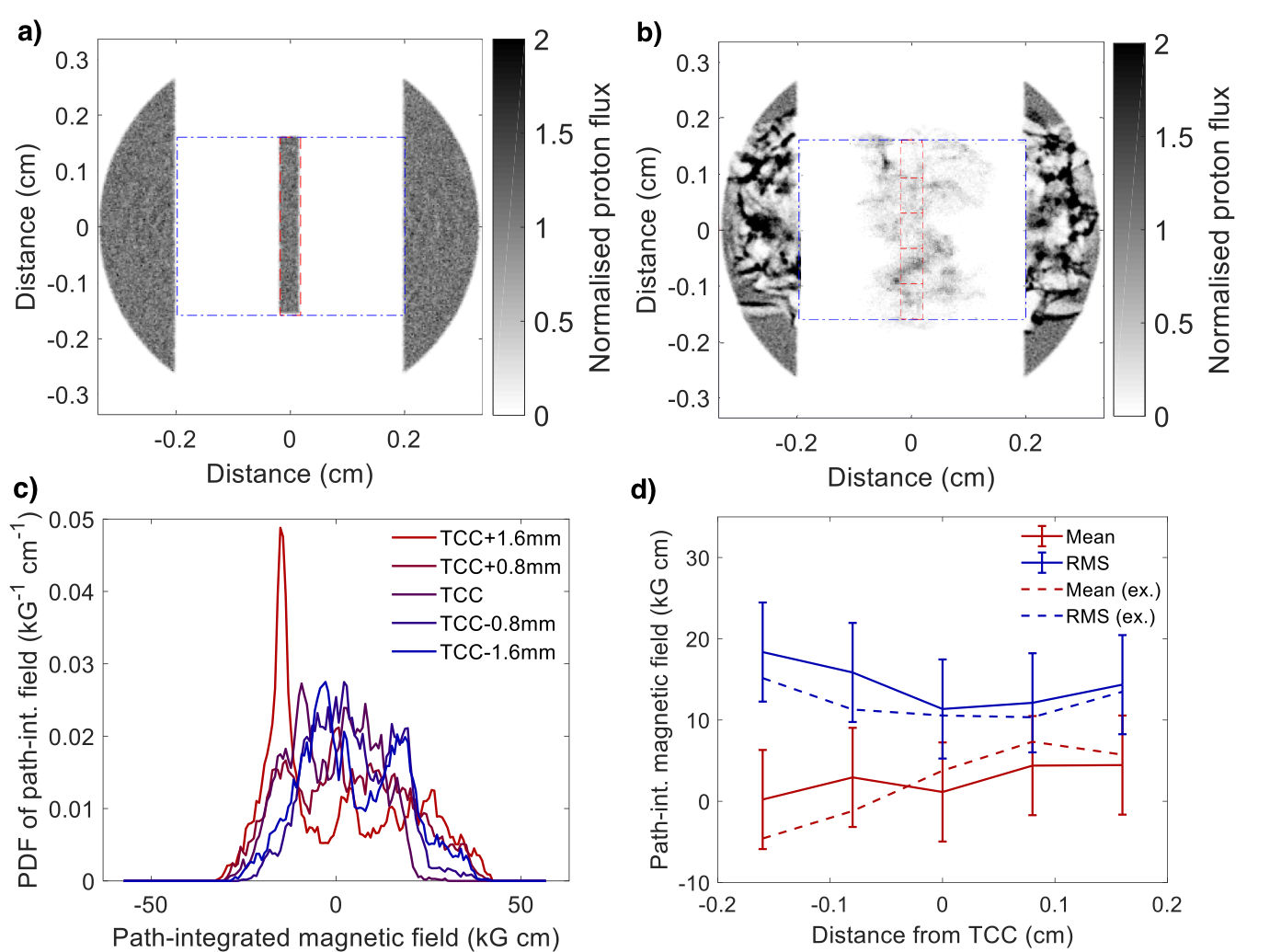}
\caption{\textit{Testing the efficacy of the slit-aperture technique using the FLASH simulations.} \textbf{a)} Initial proton-flux distribution introduced into proton-radiography module in FLASH simulations, in order to replicate effect of slit aperture on the proton beam. In the absence of any magnetic fields, the resulting proton image would reproduce this proton-flux distribution. The normalization of the proton-flux distribution is calculated using the mean proton flux over regions in which protons are not blocked by the shield. \textbf{b)} Proton-flux distribution resulting from propagating the initial proton-flux distribution shown in a) through the simulated magnetic fields arising from the FLASH simulations of the NIF experiment. The normalization for the proton-flux distribution is the same as a). \textbf{c)} Predicted distributions of the FLASH-simulated axial path-integrated fields at five distinct spatial locations along the slit. \textbf{d)} Predicted RMS and mean axial magnetic-field strengths at five spatial locations along the slit calculated from the distributions shown in c), along with the exact results calculated directly from the FLASH-simulated path-integrated axial field.}
\label{PRADfigSID}
\end{figure}

Under the assumption of isotropic magnetic fields (and assuming that the correlation length $\ell_B$ and spatial extent $\ell_n$ of those fields is known), we can relate the RMS of the axial component of the path-integrated magnetic field
to the RMS of the magnetic field itself. More specifically, it can be shown that 
\begin{equation}
   \left(\int_0^{\ell_n} \mathrm{d} s \, B_{\|}\right)_{\rm RMS}^2 = \frac{1}{2} B_{\rm RMS}^2 \ell_B \ell_n \, . \label{pathintfield}
\end{equation}
Equation (\ref{pathintfield})  is used in the main text to determine $B_{\rm RMS}$, assuming $\ell_n \approx 0.2 \, \mathrm{cm}$ (a value derived from the spatial extent of the plasma seen in the X-ray images), and $\ell_B \approx 0.01 \, \mathrm{cm}$. The lower bound on the maximum magnetic field $B_{\rm max}$ is established by noting that  the kurtosis of the path-integrated axial magnetic field is always smaller than the kurtosis of the field itself, and so
\begin{equation}
    B_{\rm max} \geq B_{\rm RMS} \frac{ \left(\int_0^{\ell_n} \mathrm{d} s \, B_{\|}\right)_{\rm max}}{ \left(\int_0^{\ell_n} \mathrm{d} s \, B_{\|}\right)_{\rm RMS}} \, . \label{maxfieldbound}
\end{equation}
The maximum displacement $\Delta \tilde{x}$ of protons is found to be $\Delta \tilde{x} \approx 2 \, \mathrm{cm}$, which in turn gives $(\int_0^{\ell_n} \mathrm{d} s \, B_{\|})_{\rm max} \approx 70 \, \mathrm{kG} \, \mathrm{cm}$. Equation (\ref{maxfieldbound}) then gives  the lower bound claimed in the main text. 

\begin{figure}[!h]
\centering
    \includegraphics[width=0.9\linewidth]{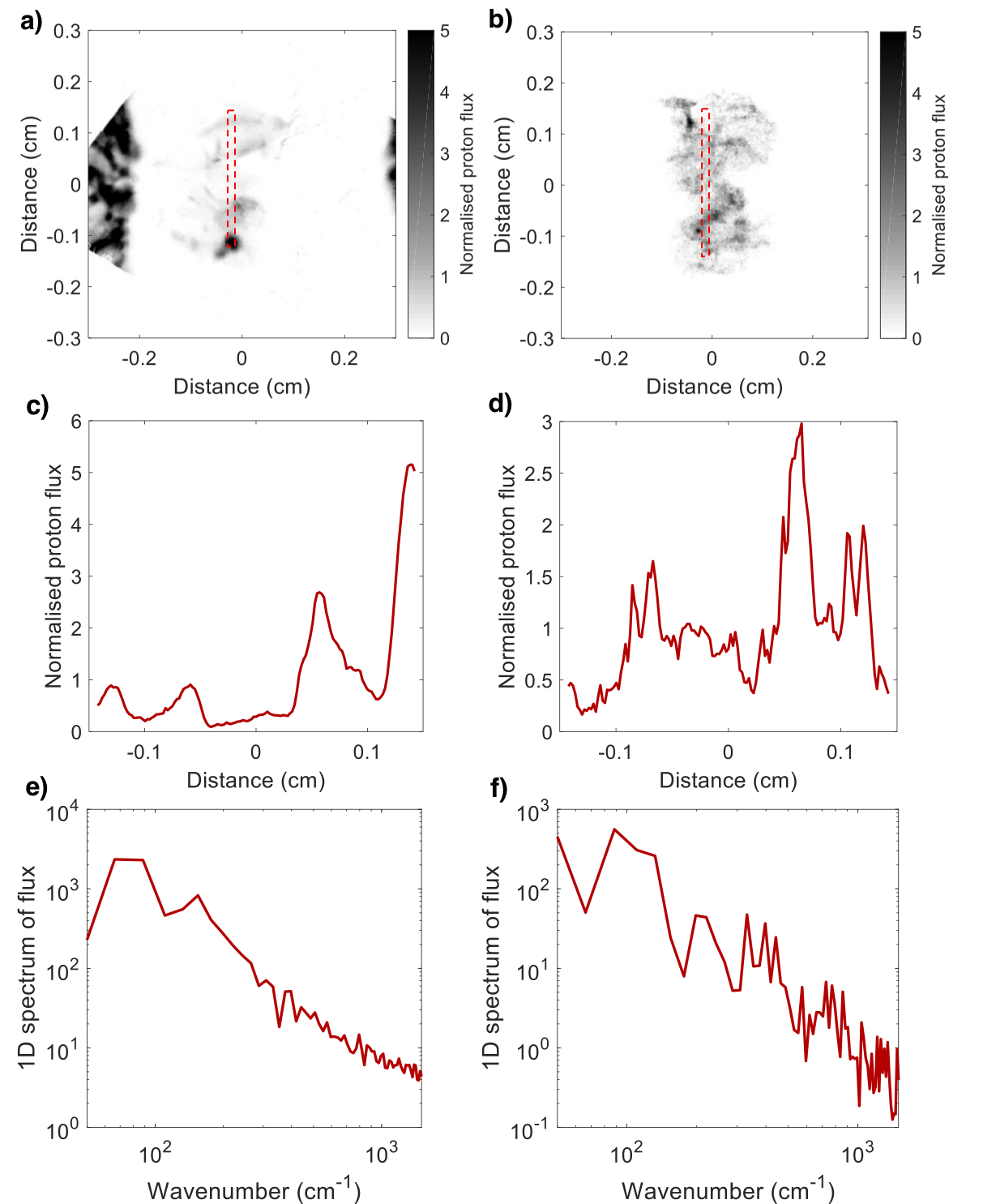}
\caption{\textit{Measuring the correlation length from the scale of flux inhomogeneities in the direction parallel to the slit aperture.} \textbf{a)} 15.0-MeV-proton image shown in Figure 3b of the main paper, rotated so that the direction parallel to the slit is vertical. The region along which averaged lineouts of the proton-flux distribution are to be taken is demarcated by a red dashed line. \textbf{b)} Simulated 15.0-MeV-proton image (with the pre-imposed slit aperture) of the FLASH-simulated magnetic fields shown in Figure \ref{PRADfigSID}b. Similarly to a), the region along which averaged lineouts of the proton-flux distribution are to be taken is demarcated by a red dashed line. \textbf{c)} Averaged lineout of the proton-flux distribution, calculated for the region indicated in a). The average is taken in the direction perpendicular to the slit orientation i.e. across the slit. \textbf{d)} As  c), but calculated using the FLASH-simulated proton image. \textbf{e)} One-dimensional spectrum of the proton-flux calculated from the averaged lineout shown in c). \textbf{f)} One-dimensional spectrum of the proton-flux calculated from the averaged lineout shown in d).}
\label{PRADfigSIH}
\end{figure}

An experimental test of the isotropic-field assumption -- as well as a validation of the results derived from the slit aperture -- was provided using a pinhole aperture to block the proton beam everywhere except for protons passing through the center of the interaction region (see Figure \ref{PRADfigSIC}).
In this set-up, the projected position of the beam in the absence of fields is clear (centered at the middle of the detector, with a ${\sim}150 \, \mu \mathrm{m}$ uncertainty in all directions), and thus Equation (\ref{PRADSI_protproj_rev}) can be used to derive the distribution of both the axial and perpendicular components of the path-integrated magnetic field. Qualitatively, we see that protons are indeed deflected in two dimensions, with displacements of similar magnitudes for both components. 

\subsection*{Testing our approach with FLASH-simulated magnetic fields}
\label{FLASH_PRAD_sims}

To validate our approach for analyzing the slit-aperture proton images described in the previous section, 
we simulate artificial slit-aperture proton images of the
magnetic fields arising in a particular FLASH simulation (25 ns, with Spitzer heat conduction switched on). We then analyze these images 
in the same manner as the experimental data; the resulting predicted values for the characteristic magnetic field 
can be compared with the true values. This analysis is presented in Figure \ref{PRADfigSID}. 

 When compared with Figure \ref{PRADfigSID}a,  Figure \ref{PRADfigSID}b shows that the megagauss magnetic fields 
arising in the FLASH simulation lead to scattering of 15.0 MeV protons passing 
through the slit aperture that is qualitatively similar to that observed 
 experimentally. As with the experimental data, we predict one-dimensional PDFs of the 
axial path-integrated magnetic field in five regions partitioning the length of 
the slit aperture (see Figure \ref{PRADfigSID}c) by assuming a correspondence between this quantity and 
one-dimensional proton-flux distributions calculated for each region by integrating the two-dimensional 
proton-flux distributions along their lengths. From these distributions, we then 
predict the RMS and mean path-integrated axial magnetic fields -- and compare 
these predictions to the true quantities calculated directly from the FLASH-simulated fields (Figure \ref{PRADfigSID}d).
We find agreement between these two quantities within the reported
uncertainty, which is calculated in the same manner as was done for the 
experimental data. 

\subsection*{Determining the correlation length of the magnetic fields}

In the main text, it was claimed that the correlation length of the magnetic 
field in the experiment was given by $\ell_B \approx 100 \, \mu \mathrm{m}$; 
this claim too can be validated with the help of FLASH simulations (see Figure \ref{PRADfigSIH}). 
If the 
characteristic deflections of protons are large enough to cause the imaging proton beam 
to self-intersect before reaching the detector, then the characteristic scale of any inhomogeneities in the proton-flux 
distribution diverges from the stochastic fields resulting in those 
inhomogeneities. However, in the FLASH simulations, we can compare
the \color{black}{correlation length $\ell_{\Psi,\mathrm{FLASH}}$ }\color{black} of inhomogeneities in the simulated slit-aperture proton-flux 
distribution with the correlation length \color{black}{$\ell_{B,\mathrm{FLASH}}$ of the magnetic field }\color{black} directly; this allows an estimate to be 
derived for $\ell_B$ obtained in the actual experiment 
from $\ell_{\Psi}$ measured experimentally \color{black}{via $\ell_B \approx \ell_{\Psi} \ell_{B,\mathrm{FLASH}}/\ell_{\Psi,\mathrm{FLASH}}$ (viz., assuming that the ratios $\ell_B/ \ell_{\Psi}$ and $\ell_{B,\mathrm{FLASH}}/\ell_{\Psi,\mathrm{FLASH}}$ are the same, though \emph{not} necessarily that $\ell_B \approx \ell_{B,\mathrm{FLASH}}$).}

We determine $\ell_{\Psi}$ and $\ell_{\Psi,\mathrm{FLASH}}$ as follows. We take lineouts 
along the direction parallel to the slit aperture; the regions used to calculate these
lineouts for the experimental and FLASH-simulated data are depicted in Figures \ref{PRADfigSIH}a and \ref{PRADfigSIH}b,
respectively, and the lineouts themselves in Figures \ref{PRADfigSIH}c and 
\ref{PRADfigSIH}d. We then determine the one-dimensional spectrum $E_{\Psi}(k)$
of each lineout (see Figures \ref{PRADfigSIH}e and \ref{PRADfigSIH}f). Finally, we 
then calculate the correlation length using the relation
\begin{equation}
  \ell_{\Psi} = \frac{1}{4 \delta \Psi_{\rm rms}^2} \int_{0}^{\infty} \mathrm{d} k \, \frac{E_{\Psi}(k)}{k} 
  \, .
\end{equation}
We find that $\ell_{\Psi} \approx \ell_{\Psi,\mathrm{FLASH}} \approx 300 \, \mu 
\mathrm{m}$. We conclude that $\ell_B$ is likely similar to $\ell_{B,\mathrm{FLASH}}$. In the case of the FLASH simulations, the correlation length can be calculated 
directly, and is found to be $\ell_{B,\mathrm{FLASH}} \approx 100 \, \mu 
\mathrm{m}$ inside the interaction region. We therefore use this value for the experiment. 

\color{black}

We note that related experiments on the Omega laser have directly measured values of $\ell_B$ whose ratio with respect to the outer scale $L$ of the turbulent cascade is (up to) ${\sim}$50\% larger than observed here ($\ell_B \approx L/4$)~\cite{bott2021time}. However, we do not believe that these results are inconsistent with those presented in this paper, because it was also shown in~\cite{bott2021time} that the technique with which the correlation length was measured can manifest systematic biases towards larger values of up to a similar (${\sim}50\%$) magnitude as the discrepancy between the Omega and NIF values of $\ell_B$. Furthermore, the uncertainty on the reported RMS magnetic-field strength $B_{\rm RMS}$ due to a ${\sim}$50\% uncertainty on $\ell_B$ is only ${\sim}25\%$; this uncertainty does not therefore affect any of our main qualitative conclusions. We also observe that FLASH simulations of the experiments on the Omega Laser Facility gave notably smaller correlation lengths ($\ell_{B,\mathrm{FLASH}} \approx L/10$) than the experimental data (in spite of modelling the hydrodynamics accurately), and also smaller values of $\ell_{\Psi}$ than we attained experimentally. There were several possible explanations that were put forward to explain the discrepancy seen in Ref.~\cite{bott2021time}; because the physical processes underpinning these explanations are also present in the NIF experiment, it remains an open question as to why exactly the FLASH simulations of the NIF experiment seem to give a similar correlation length to that attained experimentally, but the simulations of the Omega experiment do not. We will explore this question in future work. 

\color{black}

\bibliography{Refs}